\documentclass[10pt,twocolumn]{article}

\usepackage[utf8]{inputenc}
\usepackage[margin=0.75in]{geometry}
\usepackage{amsmath,amssymb,amsfonts,bm,mathtools}
\usepackage{graphicx}
\graphicspath{{../plots/}{plots/}}
\usepackage{booktabs}
\usepackage{multirow}
\usepackage{array}
\usepackage{microtype}
\usepackage{xcolor}
\usepackage{caption}
\usepackage{hyperref}
\usepackage{cleveref}

\hypersetup{
    colorlinks=true,
    linkcolor=blue!70!black,
    citecolor=green!50!black,
    urlcolor=blue!80!black
}

\renewcommand{\S}{\ensuremath{\mathsection}}
\newcommand{\R}{\mathbb{R}}
\newcommand{\E}{\mathbb{E}}
\newcommand{\keff}{k_{\mathrm{eff}}}
\newcommand{\Kstar}{K^*}
\newcommand{\norm}[1]{\left\lVert #1 \right\rVert}

\newcommand{\Fcal}{\mathcal{F}}
\newcommand{\Gcal}{\mathcal{G}}

\title{\textbf{Holo-NN: Holographic Nearest Neighbor Search via Compressive Sensing}}

\author{
    \textbf{Bryant Gipson}\\
    Google\\
    \texttt{bryantgipson@google.com}
}
\date{}

\begin{document}

\maketitle

\begin{abstract}
In the age of AI, privacy-preserving approximate nearest neighbor (ANN) search over high-dimensional embeddings has emerged as a critical yet frustratingly challenging domain, dominated by tradeoffs between speed, privacy, and accuracy. Traditional tree and graph-based indexing structures (e.g., HNSW, ScaNN) rely fundamentally on non-linear operations that are computationally intractable under Fully Homomorphic Encryption (FHE) or blind server-side constraints. 

In this paper, we introduce \textbf{Holo-NN} (\textbf{Holo}graphic \textbf{N}earest \textbf{N}eighbor Search), an algebraic framework that replaces data-dependent branching with a single linear projection. Our key observation is that while distances across a database are dense, neighborhood membership is inherently sparse: only a handful of points ever lie close to a query. By pairing locality-preserving feature transforms with Compressive Sensing, Holo-NN folds the entire database offline into a compact holographic sensing matrix. At query time, encrypted search reduces to a single, non-interactive matrix-vector multiplication---requiring no FHE bootstrapping, no multi-round interaction, and no client-side index state. This turns private retrieval into an embarrassingly parallel, accuracy-tunable computation that runs out of the box on both CPUs and GPUs, scaling efficiently to production databases. We offer a comparison against other state-of-the-art Private ANN algorithms to help situate the method in the broader context of production-grade private retrieval. 
\end{abstract}

\section{Introduction}

\subsection{Motivation \& Private Retrieval Landscape}
Published in 1750, Cramer's rule~\cite{cramer1750introduction} established an exceptionally inefficient and computationally intensive method for solving systems of linear equations based on an exhaustive calculation of determinants. And yet, despite its inefficiency, it introduced an embarrassingly parallel mathematical primitive that modern high-throughput computational hardware can execute with extreme efficiency. 

A similar paradigm shift governs modern privacy-preserving computation. The constraints imposed by fully end-to-end secure computation---such as Fully Homomorphic Encryption (FHE) and Private Information Retrieval (PIR)---fundamentally alter algorithmic tradeoffs, introducing choices that would be sub-optimal in most other settings. In unconstrained plaintext environments, Approximate Nearest Neighbor (ANN) search is dominated by hierarchical graphs (e.g., HNSW~\cite{malkov2018efficient}) and quantized inverted files (e.g., ScaNN~\cite{guo2020accelerating}), which exploit non-linear geometric branching to achieve sublinear query times ($\mathcal{O}(\log N)$ for hierarchical graphs such as HNSW). In contrast, in private, blind-server settings, data-dependent branching leaks memory access patterns, rendering graph traversals either insecure, reliant on non-colluding multi-server assumptions (e.g., PRECO~\cite{servanschreiber2022private}), or subject to heavy multi-round cryptographic overheads across garbled circuits, oblivious RAM, or sublinear PIR (e.g., SANNS~\cite{chen2020sanns}, Tiptoe~\cite{henzinger2023private}, Pacmann~\cite{zhou2024pacmann}, Panther~\cite{li2024panther}, Compass~\cite{zhu2025compass}). Under strict single-server privacy constraints, non-interactive systems are forced into an $\mathcal{O}(N)$ linear scan of the entire database.
When server compute and memory capacity are abundant relative to communication bandwidth and cryptographic circuit depth, algorithm design shifts toward strictly linear, non-interactive mathematical formulations.

\subsection{Compressed Sensing}
Beginning with Claerbout and Muir in 1973~\cite{claerbout1973robust}, it was discovered that seismological reflection data could be algorithmically improved, even with apparent ``violations'' of the Nyquist-Shannon sampling criterion~\cite{shannon1949communication}. Formalized later into the robust field of Compressive Sensing, it was shown that the supposed violation was backed by the inherent sparsity of the underlying signals~\cite{donoho2006compressed,candes2006robust}. Specifically, for a signal that admits a $k$-sparse representation where $k \ll N$, significantly fewer measurements than $N$ are required for high-fidelity reconstruction. Intuitively, reconstruction is possible because one is attempting to solve for $k$ non-zero elements in an $N$-dimensional space, which can be solved efficiently via convex $\ell_1$-optimization algorithms. Formally, given an underdetermined system of linear measurements $y = A s \in \R^M$ with $M \ll N$, an unknown $k$-sparse signal $s \in \R^N$ can be recovered by solving the convex program $\min_{s} \norm{s}_1 \text{ subject to } A s = y$ (or $\norm{A s - y}_2 \le \epsilon$ under noise), provided $M = \mathcal{O}(k \log(N/k))$.

Compressive Sensing has since enjoyed a long series of successes in Signal Processing~\cite{candes2008introduction,baraniuk2007compressive,lustig2007sparse}, though has become more of a background player in modern systems design, and it is rare to see application outside of practical acquisition systems like single-pixel optical imaging~\cite{duarte2008single}, Astronomy~\cite{bobin2008compressed}, WiFi or wireless sensor networks~\cite{bajwa2006compressive}, to name a few. 

In classical Compressive Sensing, a signal $x \in \R^N$ (e.g., an image in JPEG~\cite{wallace1991jpeg}) admits a $k$-sparse representation $s \in \R^N$ under a sparsifying basis or dictionary $F \in \R^{N \times N}$ such that $x = F s$.
Instead of acquiring all $N$ ambient samples and compressing them post hoc, compressive sensing directly acquires $M \ll N$ non-adaptive linear measurements via a pseudorandom measurement matrix $A \in \R^{M \times N}$:
\begin{equation}
    y = A x = A F s
\end{equation}
The sparse coefficient vector $s$ is then reconstructed from $y$ by applying convex $\ell_1$-regularized recovery algorithms~\cite{tibshirani1996regression,beck2009fast,tropp2007signal,daubechies2004iterative,chen2001atomic}, after which the original signal $x = F s$ can be synthesized if needed.

In our nearest-neighbor setting (Section 1.3), the paradigm is even more direct: the sparse vector $s \in \R^N$ is \emph{itself} the primary object of interest. Specifically, $s$ is the database affinity vector whose coordinates indicate the nearest neighbors of a query. Consequently, recovering $s$ directly from compressed measurements $y = A s$ immediately reveals the nearest-neighbor candidates (and their approximate distances) without requiring secondary reconstruction of an underlying signal $x$. Because $A$ is pseudorandom and the database features are known \emph{a priori}, the server's linear operator can be precomputed offline, leaving only the compact vector $y \in \R^M$ to be transmitted to the client.

\subsection{Nearest Neighbor Search as Sparse Signal Recovery}
Consider a database $T = \{t_1, t_2, \dots, t_N\} \subset \R^D$ of $N$ vectors and a query point $q \in \R^D$. For a nearest-neighbor query we are interested in the domain of distances from the query point to the points in the database, most specifically with the closest point. While the pairwise distance vector $d \in \R^N$ with entries $d_i = \mathrm{dist}(q, t_i)$ is completely dense, the corresponding $k$-nearest neighbor indicator vector $s \in \{0, 1\}^N$ is, by construction, exactly $k$-sparse. In particular, for 1-NN retrieval, $s$ is a 1-hot vector containing a single non-zero entry among $N$ coordinates.

\begin{figure}[t]
    \centering
    \includegraphics[width=\linewidth]{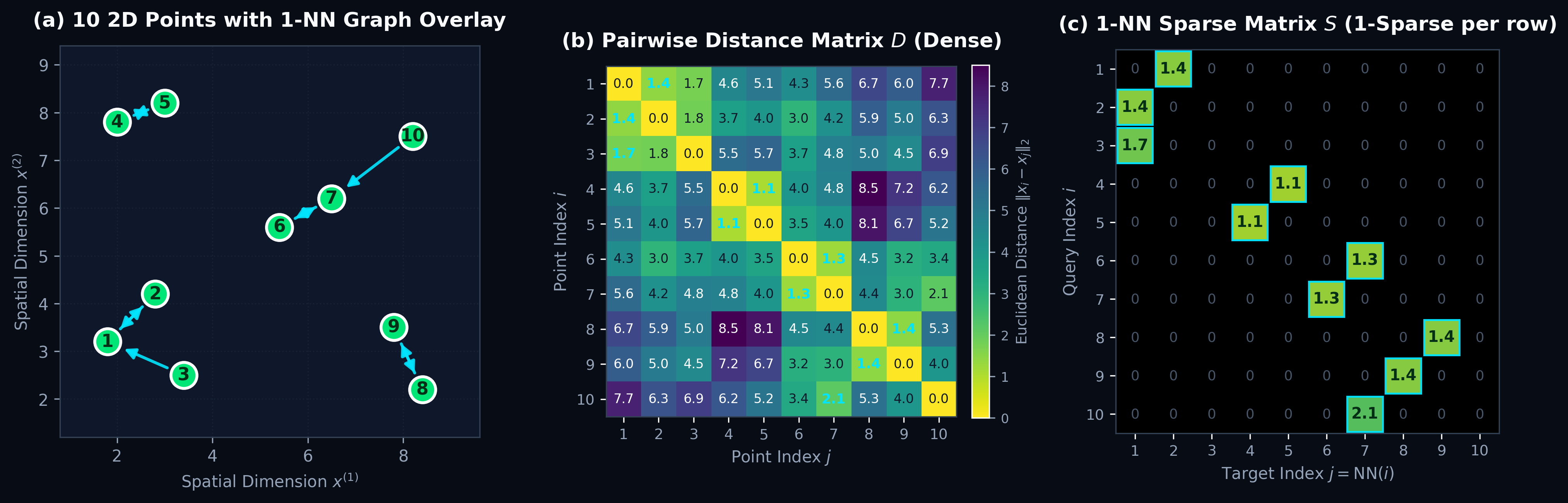}
    \caption{Nearest Neighbor search reformulated as sparse signal recovery. (a) 2D point constellation with directed 1-NN graph. (b) Dense $10 \times 10$ pairwise distance matrix $D$. (c) Distance-valued 1-sparse NN affinity matrix $\mathcal{A}$, where each row contains exactly one non-zero entry.}
    \label{fig:figure1}
\end{figure}

\begin{figure}[t]
    \centering
    \includegraphics[width=\linewidth]{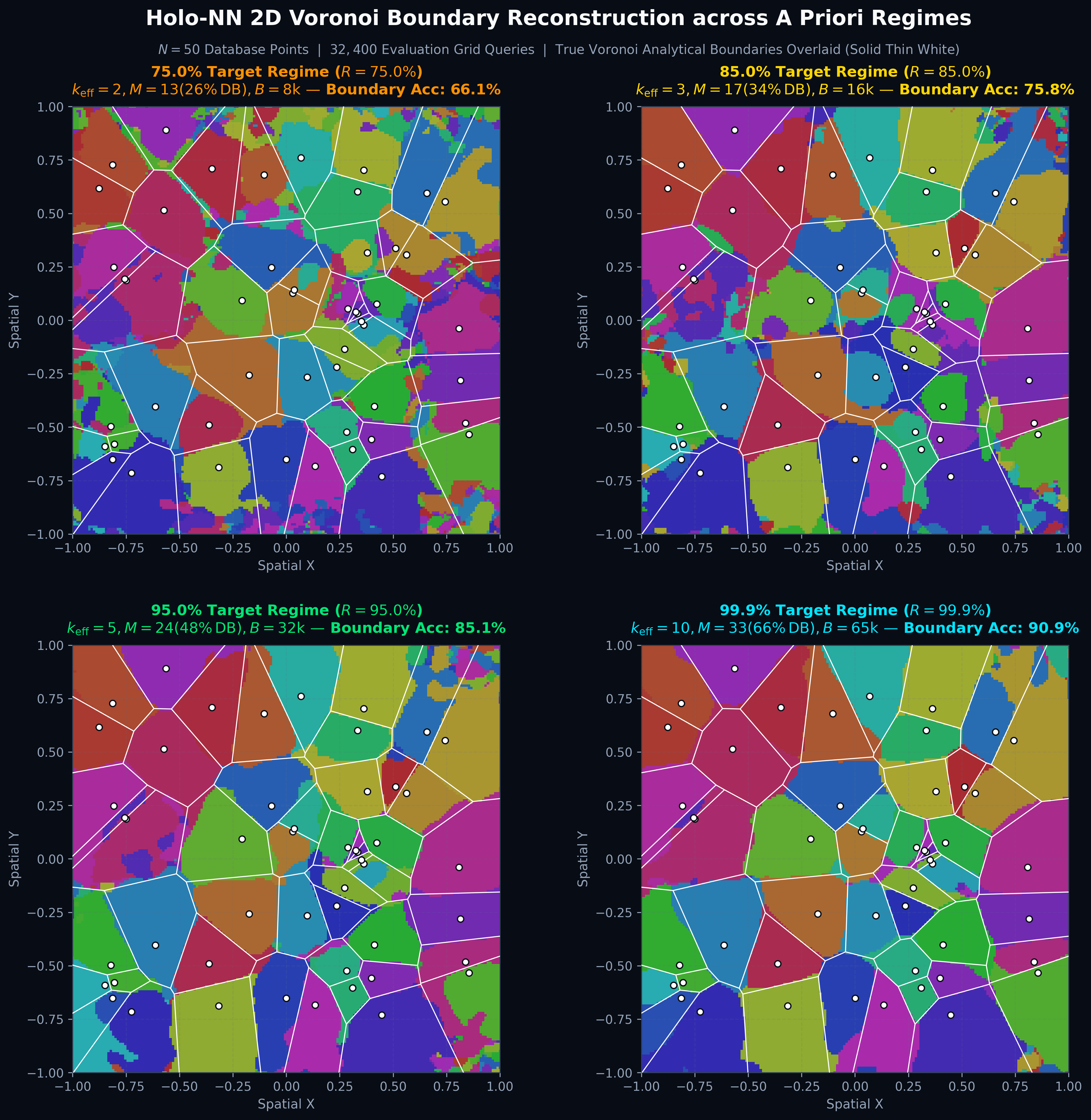}
    \caption{Holo-NN 2D Voronoi spatial boundary reconstruction across increasing measurement budgets ($75.0\%, 85.0\%, 95.0\%, 99.9\%$ recall regimes).}
    \label{fig:figure2}
\end{figure}

From this perspective, nearest neighbor search can be cast as the recovery of a sparse affinity signal from compressed linear projections. In our formulation, the target signal $s \in \R^N$ is not merely an uninformative binary 0/1 indicator, but a continuous \emph{local affinity vector} whose entries $s_i = \mathrm{sim}(q, t_i)$ concentrate their mass within the query's local Voronoi neighborhood. This continuous mass is a key feature: its relative amplitudes preserve metric resolution and local ranking among adjacent points within the cell (as visualized in Figure~\ref{fig:figure2}), while decaying rapidly to zero across distant background points to enforce effective sparsity.

The server compresses this affinity field into $M \ll N$ linear measurements $y = A s \in \R^M$ via a pseudorandom measurement matrix $A \in \R^{M \times N}$. The core challenge is designing feature embeddings $\Fcal$ and $\Gcal$ that guarantee that distances are meaningful, but that decay quickly enough such that $s$ is sufficiently sparse for stable Compressive Sensing recovery, allowing the server to compute $y = A s$ without ever computing the full $N$-dimensional vector $s$.

While the raw query $q \in \R^D$ is generally dense, and can be performed via linear operations simply (e.g., $d = u^T u + v^T v - 2 (u^T v)$ for vectors $u$ and $v$), modern machine learning provides effective non-linear feature embeddings that induce sparsity in an expanded feature space $\R^B$ ($B \gg D$). Specifically, Kernel Methods (e.g., Random Fourier Features~\cite{rahimi2007random}) and Locality Sensitive Hashing (e.g., FlyHash~\cite{dasgupta2017neural}) allow sparsified continuous metric distances to be approximated via linear inner products: 

$$\mathrm{sim}(q, t_i) \approx \langle \Gcal(t_i), \Fcal(q) \rangle$$

Where $\Gcal$ and $\Fcal$ are sparsity-enhancing feature transforms. By stacking the $N$ transformed database vectors into a database feature matrix $G = \Gcal(T) \in \R^{N \times B}$, where row $i$ is $\Gcal(t_i)^T$, the uncompressed affinity vector $s \in \R^N$ across the entire database is then simply the matrix-vector product:
\begin{equation}
    s = G \Fcal(q) = \begin{bmatrix} \langle \Gcal(t_1), \Fcal(q) \rangle \\ \vdots \\ \langle \Gcal(t_N), \Fcal(q) \rangle \end{bmatrix} \in \R^N
\end{equation}
By pre-multiplying this database feature matrix by a pseudorandom measurement matrix $A \in \R^{M \times N}$, we construct the \textbf{Holographic Sensing Matrix}:
\begin{equation}
    H \coloneqq A G \in \R^{M \times B}
\end{equation}

It's useful to pause and note the connection to Holography and physical intuition behind it. Specifically for an isolated vector pair $\langle \Gcal(t_i), \Fcal(q) \rangle$, \emph{a priori}, the computation shares no information with any neighbors and therefore the $s_i$ would need to be exhaustively calculated for a full comparison. (This is the primary mechanism behind the proof that any sorting algorithm's complexity floor is $\mathcal{O}(N \log N)$). In Holography, however, information in one basis is ``smeared out'' across a complementary basis (e.g., the Fourier Transform) ensuring that every sample contains fractional information about every point in the conjugate basis. Here $H$ plays this role, dispersing distances across multiple measurements that, collectively and in far fewer number than $N$, can infer the signal of interest. 

The linear framing cleanly separates the problem into an offline precomputation and an online linear query evaluation:
\begin{equation}
    y = H \Fcal(q) = \left(A G\right) \Fcal(q) = A \left(G \Fcal(q)\right) = A s \in \R^M
\end{equation}
Here, $H$ is computed once offline and stored permanently. Given a query $\Fcal(q)$, the evaluator simply computes the linear product $y = H \Fcal(q)$, producing the compressed measurements $y \in \R^M$ without ever materializing the uncompressed $N$-dimensional vector $s$. The querier, knowing only the pseudorandom matrix $A$, can then recover the sparse affinity vector $\hat{s}$ directly from $y = A s$ via standard convex recovery algorithms (discussed below) in a fully local environment, without ever needing to know or download the holographic database matrix $H$.

Moreover, because of the lossy nature of the initial transform, we can, with judicious design of the matrix $A$, precisely control the accuracy of the final output and more importantly, allow for an algorithmically tunable tradeoff between accuracy and cost.

By synthesizing Compressive Sensing with non-linear feature embeddings, this forms the basis of \textbf{Holo-NN}, a strictly linear 
approximate nearest neighbor retrieval framework designed for regimes where server storage and parallel compute are inexpensive relative to network bandwidth and client storage, and both query privacy and index protection are paramount.

\subsection{Summary of Contributions}
\begin{enumerate}
    \item \textbf{Holographic Retrieval Formulation}: We reformulate ANN search as a single-round-trip (1-RTT), non-interactive linear Compressive Sensing problem ($y = H \Fcal(q)$) utilizing an offline precomputed holographic matrix $H = A \Gcal(T) \in \R^{M \times B}$.
    \item \textbf{Asymmetric Neural Hashing \& Geometric Duality}: We characterize the manifold duality governing sparse feature lifting---showing why continuous Random Fourier Features sparsify spatial Euclidean manifolds while discrete hyperplane competition is required on high-dimensional semantic hyperspheres---and introduce Asymmetric FlyHash ($k_q > k_{\mathrm{db}}$) to eliminate double-quantization dropouts at zero encrypted wire overhead (Section 4 and Section~\ref{sec:benchmarks}).
    \item \textbf{Multi-Backend Encrypted Implementation \& SOTA Comparison}: We implement Holo-NN across four complementary cryptographic backends---stateless exact-integer \textbf{LinPIR (BFV)}~\cite{linpir2023} on CPU ($3.29\text{ s}$ across 4 cores or $4.39\text{ s}$ single-threaded on Wikipedia-500k, and $3.1\text{--}8.0\text{ s}$ of server compute across 6 cores on SIFT-1M, with zero server keys), depth-1 mod-switched \textbf{OpenFHE CKKS}~\cite{badawi2022openfhe} on CPU ($1.60\text{ MB}$ wire), \textbf{Phantom CUDA CKKS}~\cite{yang2024phantom} on a single NVIDIA RTX 3090 GPU ($1.68\text{ s}$ end-to-end / $1.54\text{ s}$ server on Wikipedia-500k, and $9.5\text{--}29.7\text{ s}$ on SIFT-1M), and 32-bit RNS \textbf{HEIR/Cheddar CUDA CKKS}~\cite{ali2025heir,choi2026cheddar} on a single RTX 3090 GPU ($0.79\text{ MB}$ wire in $0.134\text{ s}$ end-to-end / $50.5\text{ ms}$ server on Wikipedia-500k, and $2.10\text{--}5.37\text{ MB}$ wire in $0.26\text{--}0.82\text{ s}$ on SIFT-1M)---and benchmark against SANNS~\cite{chen2020sanns}, PRECO~\cite{servanschreiber2022private}, Tiptoe~\cite{henzinger2023private}, Pacmann~\cite{zhou2024pacmann}, Panther~\cite{li2024panther}, Compass~\cite{zhu2025compass}, and Onyx~\cite{rathee2026onyx} across all published recall conventions ($2.10\text{--}11.13\text{ MB}$ CKKS wire on SIFT-1M in $1\text{ RTT}$, $8.3\times$--$44.2\times$ below Panther's $92.9\text{ MB}$, $159\times$--$843\times$ below SANNS's $1.77\text{ GB}$, and below Compass's $12.2\text{ MB}$ over $8\text{ RTTs}$).
\end{enumerate}

\section{Background \& Problem Formulation}

\subsection{Threat Model \& Security Setting}
We consider a client-server architecture where an untrusted server $\mathcal{S}$ hosts a large-scale database $T = \{t_1, t_2, \dots, t_N\} \subset \R^D$ of $N$ high-dimensional feature vectors, and a client holds a confidential query vector $q \in \R^D$.

The objective is to retrieve the approximate nearest neighbor $q^* = \arg\min_{t_i \in T} \mathrm{dist}(q, t_i)$ under standard metric spaces---specifically Euclidean distance $\norm{q - t_i}_2$ (e.g., via RFF) or angular cosine similarity $\frac{\langle q, t_i \rangle}{\norm{q}_2 \norm{t_i}_2}$ (e.g., via FlyHash)---subject to basic privacy constraints:

\begin{enumerate}
    \item \textbf{Query \& Access-Pattern Privacy}: Under standard IND-CPA Ring-LWE encryption, the server $\mathcal{S}$ learns nothing about the client's query $q$. Crucially, unlike graph- or tree-based private search that leaks data-dependent branching paths, the server's execution trace ($y = H \Fcal(q)$) is strictly \emph{data-oblivious}: it touches every measurement channel uniformly in a single linear pass, and hence the holographic naming, eliminating memory access-pattern leakage.
    \item \textbf{Index Protection (Bounded Single-Query Leakage)}: The client never downloads the raw database $T$ or the holographic index $H$. For an honest-but-curious client issuing a query $q$, the single $M$-dimensional response $y = A G \Fcal(q) \in \R^M$ ($M \ll N$) is an underdetermined linear projection with an $(N - M)$-dimensional null space in $\R^N$, enabling sparse recovery of the $\Kstar \ll N$ local Voronoi neighbors without transmitting raw vectors $T$. Defending against an adaptive malicious client that issues $\mathcal{O}(B)$ chosen basis queries to reconstruct $H$ (or inspects ciphertext error terms) is orthogonal to our core primitive and requires standard query rate-limiting, noise flooding for circuit privacy, and zero-knowledge proofs of query well-formedness.
    \item \textbf{Operational Model}: We target deployment settings where server-side storage and high-throughput parallel compute are abundant, but client storage and network bandwidth across wide area networks (WAN) and multi-round cryptographic communication overheads are the dominant system bottlenecks.
\end{enumerate}

\subsection{Compressive Sensing \& Restricted Isometry Property}
Compressive Sensing~\cite{donoho2006compressed,candes2006robust,foucart2013mathematical} enables the stable, high probability, reconstruction of an $N$-dimensional sparse signal 
\begin{equation}
    s \in \R^N \quad \left(|\mathrm{support}(s)| \le k \ll N\right)
\end{equation}
from an underdetermined measurement vector $y \in \R^M$ with $M \ll N$ and support the number of non-zero entries:
\begin{equation}
    y = A s
\end{equation}
where $A \in \R^{M \times N}$ is a measurement operator satisfying the \textbf{Restricted Isometry Property (RIP)}~\cite{candes2008restricted,candes2005decoding} of order $k$. Specifically, there exists an isometry constant $\delta_k \in (0, 1)$ such that for all $k$-sparse vectors $s$:
\begin{equation}
    (1 - \delta_k) \norm{s}_2^2 \le \norm{A s}_2^2 \le (1 + \delta_k) \norm{s}_2^2
\end{equation}

In practice, $A$ is drawn from appropriately chosen random ensembles, most commonly:
\begin{align}
    A &\in \{-1, +1\}^{M \times N} \quad \text{(Bernoulli)} \\
    A &\sim \mathcal{N}\left(0, \frac{1}{M}\right)^{M \times N} \quad \text{(Gaussian)}
\end{align}

Though, given that Bernoulli amounts to random sums and negations of existing column entries, this distribution is typically the most efficient choice.

Because $A$ is pseudorandom, any entry $(i, j)$ or column can be evaluated on the fly in $\mathcal{O}(1)$ time by both client and server using a Pseudorandom Function (PRF) $F_k(i, j)$ keyed by a compact shared seed $k$ (e.g., 128-bit AES-CTR), eliminating any requirement to transmit or materialize $A$ in memory or GPU VRAM. 

\subsection{Kernel Lifting \& Distance Approximations}
In metric nearest neighbor search, non-linear geometric distances ($L_2$ Euclidean distance or cosine similarity) can be directly evaluated via linear operations in the raw feature space $\R^D$ ($u^T u + v^T v - 2 u^T v$), but they are typically dense in $N$. To bridge this gap, feature mapping functions $\Fcal, \Gcal: \R^D \to \R^B$ lift continuous vectors into an expanded $B$-dimensional space ($B \gg D$) where a localized similarity kernel $\mathrm{sim}(q, t_i)$---defined as a positive, monotonically decreasing function of metric distance $\mathrm{dist}(q, t_i)$ that vanishes outside a local neighborhood---is preserved as a linear inner product:
\begin{equation}
    \mathrm{sim}(q, t_i) \approx \langle \Fcal(q), \Gcal(t_i) \rangle
\end{equation}

While many kernel feature mappings exist, two are of particular use for our purposes:
\begin{enumerate}
       \item \textbf{Gaussian Radial Basis Function via Random Fourier Features}: Approximating a radial distance function, Gaussian Radial Basis Function (RBF)~\cite{schoenberg1938metric} via Random Fourier Features (RFF)~\cite{rahimi2007random} allows $L_2$ minimization over a tunable radius without the need for an interactive rounding procedure. Via Bochner's theorem~\cite{bochner1933monotone}, shift-invariant kernels such as the RBF $K(x, y) = \exp(-\gamma \norm{x - y}_2^2)$ are approximated by projecting vectors onto random Gaussian frequencies $W \sim \mathcal{N}(0, 2\gamma I)$ to produce $\mathbf{f}(x) \in \R^B$:
    \begin{equation}
        \mathbf{f}(x) = \frac{1}{\sqrt{B/2}} \begin{bmatrix} \cos(Wx) \\ \sin(Wx) \end{bmatrix}
    \end{equation}
    The RBF acts as a spatial bandpass filter: points within a Euclidean radius $\propto 1/\sqrt{\gamma}$ produce positive inner products, while distant points decay exponentially toward zero, inducing physical sparsity.
    \item \textbf{FlyHash (Biologically Inspired Sparse Hashing)}: Serving as an angular analog to the Euclidean RBF, FlyHash~\cite{dasgupta2017neural} has frequently~\cite{liang2021can,ryali2020bio} been used to approximate cosine similarity across directional semantic embeddings. Inspired by the \emph{Drosophila} olfactory circuit~\cite{caron2013random}, continuous vectors are projected onto random sparse binary hyperplanes $W \in \{0, 1\}^{B \times D}$ and can be sparsified via winner-take-all ($k$-WTA) competitive inhibition to yield a binary code, as in traditional applications, or a sparse gated projection as in our work $\mathbf{f}(x) \in \mathbb{R}^B$ whose $j$-th coordinate is:
    \begin{equation}
    \begin{aligned}
        \mathbf{f}(x)_j &= [\mathrm{WTA}_k(Wx)]_j \\
        &= \begin{cases} (Wx)_j  & \text{if } (Wx)_j \in \mathrm{Top\text{-}}k(Wx) \\ 0 & \text{otherwise} \end{cases}
    \end{aligned}
    \end{equation}
    
    Much like RBF creates localized hyperspheres in Euclidean space, FlyHash creates localized angular cones on the hypersphere, mapping angular proximity directly to the inner product overlap $\langle \mathbf{f}(x), \mathbf{f}(y) \rangle$ of sparse representations while remaining resilient to high-dimensional metric concentration.
\end{enumerate}

Throughout the remainder of this paper, the query mapping $\Fcal: \R^D \to \R^B$ and database mapping $\Gcal: \R^D \to \R^B$ denote these localized feature liftings---instantiated via RFF for Euclidean metric manifolds or Asymmetric FlyHash (Section 4) for directional semantic embeddings.

\section{The Holo-NN Retrieval Architecture}

\subsection{\texorpdfstring{Feature Projections \& Shared Feature Space ($\Fcal, \Gcal$)}{Feature Projections \& Shared Feature Space (F, G)}}
Given database points $t_i \in T$ and query $q$, the client and server apply complementary localized feature mappings $\Fcal, \Gcal: \R^D \to \R^B$ (instantiated via RFF or Asymmetric FlyHash as detailed in Section 4). These mappings are constructed to satisfy two properties: (1) \emph{Metric Locality}, preserving local neighborhood similarity $\langle \Gcal(t_i), \Fcal(q) \rangle \approx \mathrm{sim}(q, t_i)$, and (2) \emph{Zero-Mean Background Incoherence}, where global mean-centering ($\Gcal(t_i) \leftarrow \Gcal(t_i) - \mu$) ensures that background inner products vanish in expectation ($\E[\langle \Gcal(t_j), \Fcal(q) \rangle] = 0$).
Stacking the $N$ transformed database vectors produces the database embedding matrix:
\begin{equation}
    G = \Gcal(T) = \begin{bmatrix} \Gcal(t_1)^T \\ \Gcal(t_2)^T \\ \vdots \\ \Gcal(t_N)^T \end{bmatrix} \in \R^{N \times B}
\end{equation}
The uncompressed affinity vector $s \in \R^N$ between query $q$ and all $N$ database points is:
\begin{equation}
    s = G \Fcal(q) = \begin{bmatrix} \langle \Gcal(t_1), \Fcal(q) \rangle \\ \vdots \\ \langle \Gcal(t_N), \Fcal(q) \rangle \end{bmatrix} \in \R^N
\end{equation}
Because $\Fcal$ and $\Gcal$ act as localized spatial or angular bandpass filters, $s$ is \emph{$\keff$-compressible} (approximately sparse): its $\keff$ largest positive components capture the local Voronoi neighborhood around $q$, while the residual background tail decays rapidly below the universal soft-thresholding floor $\lambda$.
\subsection{Offline Server Database Grating} 
Rather than storing or scanning $G \in \R^{N \times B}$ at query time, the server pre-multiplies the database feature matrix with the pseudorandom measurement matrix $A \in \R^{M \times N}$:
\begin{equation}
    H = A G = A \Gcal(T) \in \R^{M \times B}
\end{equation}

The resulting matrix $H$ is the \textbf{Holographic Sensing Matrix}. It compresses the $N$-point database into $M$ linear measurement channels across the $B$ feature dimensions. This computation is executed strictly \textbf{offline}, once, during index construction and stored permanently on the server. Given the offline pre-computation of $H$, the function of the server benefits from theoretical or empirical optimization on algorithm parameters and hash choice. Further, because this computation is embarrassingly parallel, it can be seamlessly partitioned and executed in batch or form the basis for hierarchical design as needed.

\subsection{\texorpdfstring{Non-Interactive Linear Query Evaluation ($y = H \Fcal(q)$)}{Non-Interactive Linear Query Evaluation (y = H F(q))}}
At query time, the client computes its $B$-dimensional feature representation $\Fcal(q)$ locally and transmits it to the server (either in sparse-plaintext for bandwidth-constrained ANN, or encrypted under a Linearly Homomorphic Encryption scheme as detailed in Section 6). The server then executes a single linear matrix-vector evaluation:
\begin{equation}
    y = H \Fcal(q) \in \R^M
\end{equation}
By association:
\begin{equation}
\begin{aligned}
    y &= H \Fcal(q) = \left(A G\right) \Fcal(q) \\
      &= A \left(G \Fcal(q)\right) = A s
\end{aligned}
\end{equation}
The server thus produces the exact Compressive Sensing measurement vector $y \in \R^M$ of the latent sparse similarity signal $s \in \R^N$, without ever materializing or inspecting the full $N$-dimensional vector $s$. Crucially, because the server operator $H$ is strictly linear, cryptographic deployment does not require deep Fully Homomorphic Encryption or bootstrapping; any Linearly Homomorphic Encryption (LHE) or Ring-LWE SIMD scheme suffices to evaluate $y = H \cdot \mathrm{Enc}(\Fcal(q))$ directly over ciphertexts.

\subsection{Client Enclave Sparse Recovery \& Candidate Reranking} \label{sec:recovery}
Upon receiving the $M$-dimensional measurement vector $y$, the client recovers the sparse similarity vector $\hat{s} \in \R^N$ by solving the non-negative $\ell_1$-regularized inverse problem (LASSO~\cite{tibshirani1996regression}):
\begin{equation}
    \hat{s} = \arg\min_{s \ge 0} \frac{1}{2} \norm{A s - y}_2^2 + \lambda \norm{s}_1
\end{equation}

The client solves this convex optimization problem locally (or within a secure enclave) using any popular sparse recovery algorithm, such as Orthogonal Matching Pursuit (OMP)~\cite{tropp2007signal}, Iterative Shrinkage-Thresholding (ISTA)~\cite{daubechies2004iterative}, or the GPU-accelerated \textbf{Fast Iterative Shrinkage-Thresholding Algorithm (FISTA)}~\cite{beck2009fast} as we do in this work:
\begin{align}
    u^{(k)} &= z^{(k)} - \frac{1}{L} A^T \left(A z^{(k)} - y\right) \\
    s^{(k+1)} &= \mathcal{S}_{\lambda / L}\left(u^{(k)}\right) = \max\left(0, u^{(k)} - \frac{\lambda}{L}\right) \\
    t_{k+1} &= \frac{1 + \sqrt{1 + 4 t_k^2}}{2} \\
    z^{(k+1)} &= s^{(k+1)} + \frac{t_k - 1}{t_{k+1}} \left(s^{(k+1)} - s^{(k)}\right)
\end{align}
with initial conditions $s^{(0)} = z^{(1)} = 0, t_1 = 1$, Bai-Yin spectral step size $L = \norm{A^T A}_2 \approx (1 + \sqrt{N_{\mathrm{block}}/M_{\mathrm{block}}})^2$~\cite{yin1988limit}, and Donoho-Johnstone universal threshold $\lambda = \frac{1}{M_{\mathrm{block}}}\sqrt{2 N_{\mathrm{block}} \ln N_{\mathrm{block}} / D}$~\cite{donoho1994ideal}.

Upon convergence, the client extracts the top $\Kstar$ coordinates with the highest recovered affinity scores:
\begin{equation}
    \mathrm{Top\text{-}}\Kstar(\hat{s}) \subset \{1, 2, \dots, N\}
\end{equation}
When only candidate IDs are required (or when the client caches lightweight metadata locally), retrieval terminates here in a single round trip. When full record payloads or exact metric reranking over the $\Kstar$ candidates is required, the client issues a secondary Private Information Retrieval (Batch PIR, e.g., via HintlessPIR~\cite{linpir2023}) re-query for those $\Kstar$ indices and computes exact distances locally without revealing the candidate support to the server.

\subsection{Offline Index Tuning \& Data-Dependent Operating Curves}
In practice, the dimensions $M$ (measurement count), $B$ (feature embedding width), and $\Kstar$ (candidate verification budget) establish an operational tradeoff between accuracy, communication bandwidth, and computation. While Compressive Sensing theory establishes that $M$ scales logarithmically with ambient dimension for a fixed sparsity level ($M = \mathcal{O}(\keff \ln(\frac{N}{\keff}))$~\cite{donoho2005neighborliness,donoho2009observed}), the effective sparsity $\keff$ of the affinity vector $s = \Gcal(T)\Fcal(q)$---and the magnitude of background cross-talk across non-neighbors---is inherently \emph{data- and manifold-dependent}, varying with intrinsic dimensionality, metric concentration, and kernel bandwidth.

Crucially, because the holographic grating $H = A \Gcal(T)$ is constructed offline on the server, with the full database $T$ in hand, Holo-NN requires no brittle closed-form heuristics to guess $M$ or $\Kstar$ in the blind. Much like tuning graph degree or probe budgets (e.g., \texttt{efSearch}~\cite{malkov2018efficient}, \texttt{nprobe}~\cite{johnson2019billion,guo2020accelerating}) in traditional plaintext ANN indices, the index operator has ample offline compute during index construction to profile recall across a validation query split and select the exact $(M, B, \Kstar)$ operating point along the dataset's empirical bandwidth--accuracy Pareto curve.

\section{Asymmetric Neural Hashing \& Geometric Duality}

\subsection{The Double-Quantization Pathology in Symmetric Sparse Coding}
High-dimensional semantic vectors (such as word embeddings and neural sentence representations) reside on continuous manifolds where continuous kernel functions often suffer from high-density concentration. To create sparse anchor representations, neural hashing algorithms draw inspiration from biological olfactory circuits~\cite{dasgupta2017neural}, projecting inputs from low-dimensional projection neurons into an expanded Kenyon cell layer ($B \gg D$) and applying winner-take-all ($k$-WTA) competitive inhibition.

In conventional symmetric sparse coding, both the database points and the query vector are pruned to an identical top-$k$ active anchor budget ($k_{\mathrm{db}} = k_q \ll B$). However, symmetric hard sparsification induces a severe \textbf{double-quantization pathology}: if a salient anchor ranks just above the top-$k$ cutoff in a database vector but falls slightly below the cutoff in the query vector, their dot product vanishes completely. This sharp thresholding causes false-negative dropouts across high-dimensional semantic benchmarks.

To circumvent this limitation in compressed sensing environments, we apply an \textbf{Asymmetric FlyHash (AFH)}, breaking the symmetry between database encoding and query representation. In this approach, the functions applied to query $\Fcal$ and database $\Gcal$ are different, with the latter being significantly sparser than the former:
\begin{enumerate}
    \item \textbf{Database Grating ($k_{\mathrm{db}} \ll B$)}: The server's precomputed database feature representation $G = \Gcal(T) \in \R^{N \times B}$ maintains a strict, ultra-sparse support ($k_{\mathrm{db}} \ll B$ per row). This ensures that the offline holographic matrix $H = A G \in \R^{M \times B}$ remains compact and computationally efficient.
    \item \textbf{Query Aperture ($k_q > k_{\mathrm{db}}$)}: The client projects the query $q$ through a broader activation aperture ($k_q = \alpha_{\mathrm{aperture}} \cdot k_{\mathrm{db}}$ with $\alpha_{\mathrm{aperture}} > 1$).
\end{enumerate}

Opening the query aperture is free in bandwidth terms across both plaintext and encrypted deployments. In plaintext, transmission communicates only the non-zero coordinates of $\Fcal(q)$, so cost grows with $k_q$ from a negligible base. Under homomorphic encryption, the query is a \emph{dense SIMD ciphertext} spanning all $B$ slots whose byte size depends only on $B$ and the RNS limb count and is therefore completely independent of $k_q$---transmitting only the active coordinates is not an option, since the support of $\Fcal(q)$ must remain hidden from the server. In both regimes, the broad query aperture allows all potentially relevant semantic dimensions while the server's sharp database grating acts as a spatial filter, eliminating false-negative dropouts with no additional communication overhead.

\subsection{Geometric Duality: Continuous Euclidean RFF vs. Discrete Hyperplane Competition}
Empirical and theoretical investigations reveal idiosyncratic, data-dependent structures governing feature transform selection. That is, while arguments for \emph{a priori} parameter selection can be made, the dominating influences on the parameter choices for a dataset come from the structure, distribution and density of the data manifold, in addition to its ambient dimension $D$. MNIST-784 ($D = 784$) falls squarely in the first regime while GloVe-25 ($D = 25$) falls squarely in the second, as $D$ alone does not predict which transform applies.
\begin{enumerate}
    \item \textbf{Spatially Structured Metric Manifolds (low intrinsic dimension)}:
    In Euclidean metric spaces with physical or spatial structure (e.g., 2D spatial seeds, MNIST-784 pixel spaces, and SIFT-128 image descriptors), pairwise Euclidean distances show meaningful local contrast. In this regime, Random Fourier Features (RFF) with Gaussian RBF kernels $K(x, y) = \exp(-\gamma \norm{x - y}_2^2)$ create genuine physical sparsity because spatial volume scales as $\mathcal{O}(r^D)$. Points outside the local kernel radius $r > 3/\sqrt{2\gamma}$ decay exponentially toward zero, enabling high-fidelity metric nearest neighbor recovery (Figure~\ref{fig:figure2} and Table~\ref{tab:table1}).
    \item \textbf{Unstructured Semantic Hyperspheres (intrinsic dimension $\approx$ ambient $D$)}:
    On unconstrained hyperspherical manifolds, pairwise Euclidean distances concentrate sharply around $\sqrt{2}$ as $D$ grows, since $\norm{x-y}_2^2 = 2 - 2\langle x, y\rangle$ and $\langle x, y\rangle \to 0$ by measure concentration on the sphere. Continuous exponential kernels suffer from amplitude collapse, where the true neighbor similarity signal drops below the finite-sample RFF Monte Carlo noise floor $\mathcal{O}(1/\sqrt{B})$.
    
    In contrast, Asymmetric FlyHash partitions high-dimensional space via nearly orthogonal random hyperplanes. Discrete $k$-WTA competition creates implicit Voronoi cells, preserving semantic locality without exponential amplitude attenuation and remaining robust to the curse of dimensionality. Moreover, because feature strength is preserved within each Voronoi cell, a local ``in-cell Euclidean metric'' breaks ties and further increases NN accuracy. 
\end{enumerate}

\section{Algorithmic ANN Benchmark Evaluation}
\label{sec:benchmarks}

\subsection{Benchmark Datasets \& Experimental Setup}
To validate the algorithmic accuracy and theoretical predictions of Holo-NN independently of cryptographic backends, we evaluate Holo-NN across four standard nearest-neighbor benchmark suites spanning Euclidean ($L_2$) and angular (cosine) metric spaces:
\begin{enumerate}
    \item \textbf{SIFT-1M ($N = 1{,}000{,}000, D = 128$)}: The canonical computer vision benchmark evaluated under \textbf{Euclidean ($L_2$) distance} across the first $Q = 1{,}000$ of the $10{,}000$ official test queries. Vectors are mean-centered, globally rescaled, and lifted via Random Fourier Features ($B = 16{,}384$, $\gamma = 8.0$), mapping Euclidean proximity to real-valued inner products.
    \item \textbf{MNIST-784 ($N = 60{,}000, D = 784$)}: Handwritten digit pixel vectors evaluated under \textbf{Euclidean ($L_2$) distance} across $Q = 10{,}000$ test queries, lifted via mean-centered Random Fourier Features ($B = 8{,}192$, $\gamma = 2.0$).
    \item \textbf{GloVe-25 ($N = 1{,}183{,}514, D = 25$)}: Web-crawl word embeddings evaluated under \textbf{angular cosine similarity} across $Q = 1{,}000$ test queries, lifted via Asymmetric FlyHash ($B = 65{,}536$) to resolve angular crowding on the unit hypersphere.
    \item \textbf{Last.fm-64 ($N = 292{,}385, D = 65$)}: Sparse user-music interaction preference vectors evaluated under \textbf{maximum inner product (MIPS)} across $Q = 1{,}000$ test queries. Because the inner product is already linear, we apply no feature lifting ($\Fcal = \Gcal = \mathrm{id}$, $B = D = 65$) and operate directly on the raw vectors. This dataset therefore has no sparsification mechanism, and serves as our boundary case.
\end{enumerate}

All benchmarks are evaluated on an NVIDIA GeForce RTX 3090 GPU ($24\text{ GB}$ VRAM) and an Intel Xeon W-2135 host ($6$ cores / $12$ threads) across four representative operating points ($75.0\%, 85.0\%, 95.0\%, 99.9\%$ nominal recall tiers). At each tier, $(M, \Kstar)$ represents the empirically profiled Pareto-optimal measurement budget and candidate cutoff along the dataset's bandwidth--recall frontier. To prevent client-side FISTA from materializing a monolithic $M \times N$ sensing matrix (which would require $1.52\text{ TB}$ of float32 memory at $N = 10^6, M = 380{,}100$), million-vector corpora are partitioned into $P = 100$ disjoint blocks of $N_{\mathrm{block}} = 10{,}000$ vectors ($M_{\mathrm{block}} = M / P$). Each block's sensing matrix $A_b \in \R^{M_{\mathrm{block}} \times N_{\mathrm{block}}}$ is drawn from the normalized Bernoulli ensemble $\pm 1/\sqrt{M_{\mathrm{block}}}$ and generated on the fly from a shared 128-bit PRNG seed ($\le 152\text{ MB}$ in GPU VRAM, fitting even on lightweight $4\text{ GB}$ client GPUs), enabling batched GPU FISTA across all $P$ blocks in $< 0.4\text{ s}$ per query.

\subsection{Multi-Scale Target Recall Benchmark Results}

\begin{table*}[t] 
\centering
\small
\setlength{\tabcolsep}{4pt}
\caption{Multi-Scale Benchmark Results across 4 Datasets and 4 Operating Tiers ($M, \Kstar$). Rec@$k$ is 1-recall@$k$: the fraction of queries whose true nearest neighbor appears among the top $k$ candidates returned by FISTA.}
\label{tab:table1}
\resizebox{\textwidth}{!}{%
\begin{tabular}{llcccccccccc}
\toprule
\textbf{Dataset} & \textbf{Metric Space \& Transform} & $N$ & $D$ & $B$ & \textbf{Regime} & $M$ & $M/N$ & $\Kstar$ & \textbf{Rec@1} & \textbf{Rec@$\Kstar$} & \textbf{Rec@50} \\
\midrule
\multirow{4}{*}{\textbf{SIFT-1M}} & \multirow{4}{*}{Euclidean RFF} & \multirow{4}{*}{1,000,000} & \multirow{4}{*}{128} & \multirow{4}{*}{16,384} 
  & \textbf{75.0\%} & 380,100 & 38.01\% & \textbf{8} & 62.00\% & \textbf{94.10\%} & 99.20\% \\
  & & & & & \textbf{85.0\%} & 445,200 & 44.52\% & \textbf{11} & 67.60\% & \textbf{98.00\%} & 99.80\% \\
  & & & & & \textbf{95.0\%} & 631,200 & 63.12\% & \textbf{19} & 72.30\% & \textbf{99.70\%} & 100.00\% \\
  & & & & & \textbf{99.9\%} & 1,294,900 & 129.49\% & \textbf{50} & 82.10\% & \textbf{100.00\%} & 100.00\% \\
\midrule
\multirow{4}{*}{\textbf{MNIST-784}} & \multirow{4}{*}{Pixel Manifold RFF} & \multirow{4}{*}{60,000} & \multirow{4}{*}{784} & \multirow{4}{*}{8,192} 
  & \textbf{75.0\%} & 6,000 & 10.00\% & \textbf{30} & 23.37\% & \textbf{75.42\%} & 79.95\% \\
  & & & & & \textbf{85.0\%} & 9,000 & 15.00\% & \textbf{30} & 32.22\% & \textbf{86.02\%} & 90.54\% \\
  & & & & & \textbf{95.0\%} & 12,000 & 20.00\% & \textbf{40} & 39.14\% & \textbf{93.80\%} & 94.95\% \\
  & & & & & \textbf{99.9\%} & 16,000 & 26.67\% & \textbf{50} & 46.46\% & \textbf{97.79\%} & 97.79\% \\
\midrule
\multirow{4}{*}{\textbf{GloVe-25}} & \multirow{4}{*}{Angular FlyHash (AFH)} & \multirow{4}{*}{1,183,514} & \multirow{4}{*}{25} & \multirow{4}{*}{65,536} 
  & \textbf{75.0\%} & 95,000 & 8.03\% & \textbf{500} & 18.00\% & \textbf{82.60\%} & 63.30\% \\
  & & & & & \textbf{85.0\%} & 130,000 & 10.98\% & \textbf{500} & 20.60\% & \textbf{89.00\%} & 70.20\% \\
  & & & & & \textbf{95.0\%} & 200,000 & 16.90\% & \textbf{500} & 24.90\% & \textbf{94.80\%} & 78.70\% \\
  & & & & & \textbf{99.9\%} & 300,000 & 25.35\% & \textbf{500} & 26.60\% & \textbf{97.50\%} & 84.30\% \\
\midrule
\multirow{4}{*}{\textbf{Last.fm-64}} & \multirow{4}{*}{Raw MIPS (no lifting)} & \multirow{4}{*}{292,385} & \multirow{4}{*}{65} & \multirow{4}{*}{65} 
  & \textbf{75.0\%} & 20,000 & 6.84\% & \textbf{50} & 6.60\% & \textbf{57.60\%} & 57.60\% \\
  & & & & & \textbf{85.0\%} & 28,000 & 9.58\% & \textbf{50} & 8.90\% & \textbf{71.10\%} & 71.10\% \\
  & & & & & \textbf{95.0\%} & 40,000 & 13.68\% & \textbf{50} & 14.40\% & \textbf{82.30\%} & 82.30\% \\
  & & & & & \textbf{99.9\%} & 60,000 & 20.52\% & \textbf{50} & 22.30\% & \textbf{93.30\%} & 93.30\% \\
\bottomrule
\end{tabular}%
}
\end{table*}

Table~\ref{tab:table1} illustrates how achieved recall tracks the sparsity induced by the feature transform across each data manifold. On spatially structured Euclidean manifolds (\textbf{SIFT-1M} and \textbf{MNIST-784}), Random Fourier Features concentrate affinity within a tight local radius, reaching $94.10\%$--$100.00\%$ on SIFT-1M and $75.42\%$--$97.79\%$ on MNIST-784 at $10.0\%$--$26.7\%$ sub-Nyquist compression (noting that pushing unclustered SIFT-1M to $100.0\%$ crosses the Nyquist boundary $M/N > 100\%$, whereas practical encrypted deployments operate at $M/N \le 50\%$ where 1-recall@10 already reaches $98.5\%$, as swept in Table~\ref{tab:table4}). On the angular semantic hypersphere (\textbf{GloVe-25}), Asymmetric FlyHash resolves crowding to achieve $82.60\%$--$97.50\%$ at $M/N = 8.03\%$--$25.35\%$. By contrast, \textbf{Last.fm-64} ($\Fcal = \Gcal = \mathrm{id}$, no feature lifting) provides no non-linear sparsification mechanism, causing background cross-talk to degrade recovery ($57.60\%$--$93.30\%$) and marking the boundary of the method's applicability.

\section{Encrypted Retrieval System Implementation \& SOTA Comparison}

Holo-NN is designed for blind cloud retrieval across four complementary Ring-LWE encrypted execution backends: (1) exact-integer BFV homomorphic matrix-vector evaluation via Google's \textbf{HintlessPIR / LinPIR} engine~\cite{linpir2023}, (2) depth-1 modulus-switched approximate homomorphic encryption on CPU via \textbf{OpenFHE}~\cite{badawi2022openfhe} under the Cheon-Kim-Kim-Song (CKKS) scheme~\cite{cheon2017homomorphic} compiled through \textbf{HEIR}~\cite{ali2025heir}, (3) 64-bit RNS GPU-accelerated CKKS evaluation on a single NVIDIA GeForce RTX 3090 GPU ($24\text{ GB}$ VRAM) via \textbf{Phantom}~\cite{yang2024phantom} with a fused 128-bit lazy-reduction BSGS inner-product kernel, and (4) 32-bit RNS double-hoisted GPU CKKS evaluation on a single RTX 3090 GPU (and NVIDIA H100) lowered via \textbf{HEIR}~\cite{ali2025heir} to \textbf{Cheddar}~\cite{choi2026cheddar}.

The encrypted retrieval pipeline executes via a strictly linear four-step protocol:
\begin{enumerate}
    \item \textbf{Offline Precomputation (Server)}: The server computes the holographic index $H = A G = A \Gcal(T) \in \R^{M \times B}$ once across all database embeddings and pre-encodes its diagonals in NTT evaluation form.
    \item \textbf{Query Encryption (Client)}: The client lifts query $q \to \Fcal(q)$ and encrypts $\mathrm{Enc}(\Fcal(q))$ into SIMD ciphertext slots.
    \item \textbf{Blind Linear Evaluation (Server)}: The server evaluates $\mathrm{Enc}(y) = H \cdot \mathrm{Enc}(\Fcal(q))$ via diagonal NTT plaintext-ciphertext multiplication with zero bootstrapping---using single-generator cyclic query rotations and rotation-free row accumulation under LinPIR, or Baby-Step Giant-Step (BSGS) evaluation followed by single-limb modulus switching (\texttt{Compress}) under CKKS.
    \item \textbf{Client Recovery \& Optional PIR Reranking}: The client decrypts $y = \mathrm{Dec}(\mathrm{Enc}(y))$ locally, reconstructs sparse coefficients $\hat{s} = \mathrm{FISTA}(y, A, \lambda, L)$ and extracts the top $\Kstar$ candidate IDs (issuing an optional secondary Batch PIR re-query for those $\Kstar$ records when exact-distance reranking or document payloads are needed; Section~\ref{sec:recovery}).
\end{enumerate}

\subsection{Degree-1 Blind Linear Arithmetic \& Compiler Optimizations}
Prior FHE-based vector search systems evaluate non-linear Euclidean distances or polynomial approximations of cosine similarity under homomorphic encryption. Evaluating polynomial approximations of degree $d \ge 12$ requires deep homomorphic circuits, inducing large noise growth that necessitates expensive \textbf{FHE bootstrapping} operations (often requiring $10$--$60\text{ seconds}$ per ciphertext).

In Holo-NN, the server's blind evaluation is strictly linear:
\begin{equation}
    \mathrm{Enc}(y) = H \cdot \mathrm{Enc}(\Fcal(q))
\end{equation}
Because $H$ is plaintext and $\mathrm{Enc}(\Fcal(q))$ is a degree-1 ciphertext, the server executes only homomorphic plaintext-ciphertext multiplications, additions, and slot rotations:
\begin{itemize}
    \item \textbf{Multiplicative Depth}: Exactly \textbf{1} (Degree-1 linear operator).
    \item \textbf{FHE Bootstrapping}: \textbf{0 bootstrapping operations} required throughout the entire query lifecycle.
    \item \textbf{Compiler Optimization via HEIR}: Lowering the matrix-vector multiplication through the MLIR-based HEIR compiler~\cite{ali2025heir} automatically synthesizes diagonal ciphertext rotations and hoists plaintext weights, yielding optimal SIMD slot packing across OpenFHE backends.
    \item \textbf{Noise Growth}: Minimal invariant noise budget, permitting compact RLWE parameter sets at OpenFHE's \texttt{HEStd\_128\_classic} security standard~\cite{albrecht2018homomorphic} ($N_{\mathrm{ring}} = 16{,}384$ for $8{,}192$-slot blocks and $N_{\mathrm{ring}} = 32{,}768$ for $16{,}384$-slot blocks).
\end{itemize}

\begin{table}[t]
\centering
\caption{Deployed encrypted Wikipedia-500k demonstrator ($N = 500{,}000$, $B = 8{,}192$, $M = 32{,}768$, $4$ blocks of $8{,}192 \times 8{,}192$). Latency is end-to-end wall clock; wire cost is total serialized request plus response ciphertexts.}
\label{tab:table2}
\small
\setlength{\tabcolsep}{3.2pt}
\resizebox{\columnwidth}{!}{%
\begin{tabular}{@{}lrrr@{}}
\toprule
\textbf{Execution surface} & \textbf{Wire / query} & \textbf{Latency} & \textbf{Server keys} \\
\midrule
Plaintext (no encryption) & $0.16\text{ MB}$ & $1.14\text{ s}$ & $0\text{ MB}$ \\
\textbf{Holo-NN: LinPIR} (CPU, BFV~\cite{linpir2023} exact-int.) & $\mathbf{31.41\text{ MB}}$ & $\mathbf{3.29\text{ s}}$ ($4.39\text{ s}^{\S}$) & $\mathbf{0\text{ MB}}$ \\
\textbf{Holo-NN: CKKS-CPU} (OpenFHE~\cite{badawi2022openfhe}, 12-thr) & $1.60\text{ MB}$ & $32.9\text{ s}$ & $299\text{ MB}^{\star}$ \\
\textbf{Holo-NN: CKKS-GPU} (Phantom~\cite{yang2024phantom}, $1\times$ RTX 3090) & $\mathbf{1.60\text{ MB}}$ & $\mathbf{1.68\text{ s}}^{\diamond}$ & $85\text{ MB}^{\star}$ \\
\textbf{Holo-NN: HEIR-GPU} (Cheddar~\cite{ali2025heir,choi2026cheddar}, $1\times$ RTX 3090) & $\mathbf{0.79\text{ MB}}$ & $\mathbf{0.134\text{ s}}^{\ddagger}$ & $\mathbf{75\text{ MB}}^{\star}$ \\
\bottomrule
\end{tabular}%
}
\vspace{2pt}
{\footnotesize $^{\S}$ First figure ($3.29\text{ s}$ end-to-end; $1.78\text{ s}$ server compute) is sharded across $4$ CPU cores ($4\times$ $8{,}192 \times 8{,}192$ shards in parallel on the 6-core Xeon W-2135); parenthetical figure ($4.39\text{ s}$ end-to-end; $2.88\text{ s}$ server compute) is single-threaded on $1$ core.
$^{\diamond}$ On $1\times$ RTX 3090 ($24\text{ GB}$ VRAM), Phantom holds all $4$ 64-bit RNS blocks ($12.0\text{ GiB}$ plaintexts, $17.1\text{ GiB}$ total VRAM) and evaluates all $4$ blocks in $\mathbf{1.54\text{ s}}$ of server compute ($\mathbf{1.68\text{ s}}$ end-to-end across $10$ queries, or $0.385\text{ s}$/block; vs.\ $6.95\text{ s}$ across $2\times$ RTX 2080 Ti GPUs).
$^{\ddagger}$ HEIR/Cheddar~\cite{ali2025heir,choi2026cheddar} uses 32-bit RNS words (\texttt{uint32\_t}, $\log_2(QP) \approx 91\text{ bits}$), halving wire footprint to $\mathbf{0.79\text{ MB}}$ ($256\text{ KiB}$ query $+ 4 \times 128\text{ KiB}$ responses) and reducing plaintext VRAM by $3\times$ to $4.00\text{ GiB}$ ($8.1\text{ GiB}$ total VRAM). It evaluates a single $8{,}192 \times 8{,}192$ block in $\mathbf{12.58\text{ ms}}$ on RTX 3090 ($\mathbf{5.33\text{ ms}}$ on NVIDIA H100; $8.67\text{ ms}$ RTX 3090 / $2.51\text{ ms}$ H100 at $4{,}096 \times 4{,}096$), completing all $4$ Wikipedia blocks in $\mathbf{50.5\text{ ms}}$ of server compute and $\mathbf{0.134\text{ s}}$ end-to-end ($100\%$ Top-1 parity with plaintext).
$^{\star}$ At $N_{\mathrm{ring}} = 16{,}384$, OpenFHE stores $190$ 64-bit RNS BSGS rotation keys ($299.0\text{ MB}$), Phantom stores $27$ NAF/dyadic keys ($84.9\text{ MB}$), and Cheddar stores $190$ 32-bit RNS hoisted BSGS keys ($74.7\text{ MB}$).\par}
\end{table}

\begin{table*}[t]
\centering
\caption{State-of-the-art comparison across private ANN retrieval architectures. System figures are transcribed from each primary publication at its reported database scale $N$. \textbf{Wire / query} is total mandatory per-query traffic (upload $+$ download). \textbf{Client / Server State} lists persistent client RAM, per-session cold-start traffic, or server-side evaluation keys. Under \textbf{Retrieval Quality}, each metric occupies its own column: \textbf{MRR@10} (Compass convention; $^{\ast}$denotes MRR@100 on MS MARCO), \textbf{1-r@10} (true 1-NN in top 10), \textbf{10-r@10} (Panther/Onyx convention: $\lvert\text{ours}_{1:10}\cap\text{true}_{1:10}\rvert/10$), and \textbf{10-r@50} (1-RTT candidate-set coverage of true 10-NN in top 50, which equals final reranked 10-r@10 after the optional $+1$ RTT Batch PIR fetch).}
\label{tab:table3}
\footnotesize
\setlength{\tabcolsep}{3.5pt}
\resizebox{\textwidth}{!}{%
\begin{tabular}{@{}llcrrrlrrrr@{}}
\toprule
& & & & & & & \multicolumn{4}{c}{\textbf{Retrieval Quality}} \\
\cmidrule(lr){8-11}
\textbf{System} & \textbf{Mechanism \& Hardware} & \textbf{RTT} & \textbf{Wire / query} & \textbf{Client / Server State} & \textbf{Latency} & \textbf{Benchmark ($N$)} & \textbf{MRR@10} & \textbf{1-r@10} & \textbf{10-r@10} & \textbf{10-r@50} \\
\midrule
\textbf{SANNS} \emph{(USENIX Sec '20)}~\cite{chen2020sanns} & AHE + DORAM + GC 2PC (\emph{server-private} DB, 32-thr) & Multi$^{\dagger}$ & $1.77\text{ GB}^{\dagger}$ & no offline hints & $1.55\text{--}45.3\text{ s}$ & \textbf{SIFT-1M} ($1\mathrm{M}$) & --- & --- & $\approx 90\%$ & --- \\
\textbf{PRECO} \emph{(IEEE S\&P '22)}~\cite{servanschreiber2022private} & LSH + DPF (\emph{2 non-colluding servers}, 32-thr) & $1^{\S}$ & $1.5\text{ MB}^{\S}$ & 2-srv DB replica & $1.1\text{--}28.2\text{ s}^{\S}$ & \textbf{SIFT-1M} ($1\mathrm{M}$) & --- & --- & $\approx 95\%$ & --- \\
\textbf{Tiptoe} \emph{(SOSP '23)}~\cite{henzinger2023private} & LHE + PIR (\emph{public} DB, 45-srv CPU) & $2^{\dagger}$ & $59.7\text{ MB}^{\dagger}$ & $0.29\text{ GiB}$ model+centroids & $2.7\text{ s}$ & Common Crawl ($364\mathrm{M}$) & ---$^{\ast}$ & --- & --- & --- \\
\textbf{Pacmann} \emph{(ICLR '25)}~\cite{zhou2024pacmann} & Sublinear PIR (\emph{public} DB, CPU) & $20$ & $61.6\text{ MB}^{\dagger}$ & $0.6\text{ GB}$ client ($2.7\text{ GB}$ init) & $1.1\text{ s}$ & MS MARCO ($3.2\mathrm{M}$) & $0.266^{\ast}$ & --- & --- & --- \\
\textbf{Pacmann} \emph{(ICLR '25)}~\cite{zhou2024pacmann} & Sublinear PIR (\emph{public} DB, CPU) & $32$ & $413.8\text{ MB}^{\dagger}$ & $2.9\text{ GB}$ client ($59.6\text{ GB}$ init) & $1.6\text{--}3.0\text{ s}$ & SIFT ($100\mathrm{M}$) & --- & --- & $\approx 90\%$ & --- \\
\textbf{Panther} \emph{(CCS '25)}~\cite{li2024panther} & LHE + Batch PIR + SS/GC 2PC (\emph{server-private} DB, 32-thr) & Multi$^{\dagger}$ & $92.9\text{ MB}^{\dagger}$ & no offline hints & $1.49\text{--}9.32\text{ s}$ & \textbf{SIFT-1M} ($1\mathrm{M}$) & --- & --- & $88.85\%$ & --- \\
\textbf{Compass} \emph{(OSDI '25)}~\cite{zhu2025compass} & Ring ORAM over HNSW (CPU) & $8$ & $12.2\text{ MB}^{\dagger}$ & $35.8\text{ MB}$ client RAM ($36.95\text{ MB}$ init) & $0.04\text{--}0.85\text{ s}$ & \textbf{SIFT-1M} ($1\mathrm{M}$) & matches BF & --- & --- & --- \\
\textbf{Onyx} \emph{(arXiv '26)}~\cite{rathee2026onyx} & Hardware TEE (SEV-SNP + ORAM) & $1^{\ddagger}$ & n/a$^{\ddagger}$ & $170\text{ MB}$ enclave ($1.5\text{ GB}$ SSD) & $0.019\text{ s}^{\ddagger}$ & \textbf{SIFT-1M} ($1\mathrm{M}$) & --- & --- & $\approx 90\%$ & --- \\
\midrule
\textbf{Holo-NN: LinPIR} ($M = 98{,}300$) & BFV exact-int.\ RLWE (6-core CPU)~\cite{linpir2023} & $\mathbf{1}\text{ (}+1\text{)}$ & $\mathbf{91.1\text{ MB}}$ & $\mathbf{0\text{ MB}}$ \textbf{client} / $\mathbf{0\text{ MB}}$ \textbf{server} & $\mathbf{3.1\text{ s}}^{\|}$ & \textbf{SIFT-1M} ($1\mathrm{M}$) & $0.417$ & $63.4\%$ & $33.3\%$ & $57.0\%$ \\
\textbf{Holo-NN: LinPIR} ($M = 300{,}000$) & BFV exact-int.\ RLWE (6-core CPU)~\cite{linpir2023} & $\mathbf{1}\text{ (}+1\text{)}$ & $\mathbf{265.4\text{ MB}}$ & $\mathbf{0\text{ MB}}$ \textbf{client} / $\mathbf{0\text{ MB}}$ \textbf{server} & $\mathbf{6.6\text{ s}}^{\|}$ & \textbf{SIFT-1M} ($1\mathrm{M}$) & $\mathbf{0.702}$ & $\mathbf{93.0\%}$ & $\mathbf{61.6\%}$ & $\mathbf{92.0\%}$ \\
\textbf{Holo-NN: LinPIR} ($M = 380{,}100$) & BFV exact-int.\ RLWE (6-core CPU)~\cite{linpir2023} & $\mathbf{1}\text{ (}+1\text{)}$ & $\mathbf{334.6\text{ MB}}$ & $\mathbf{0\text{ MB}}$ \textbf{client} / $\mathbf{0\text{ MB}}$ \textbf{server} & $\mathbf{8.0\text{ s}}^{\|}$ & \textbf{SIFT-1M} ($1\mathrm{M}$) & $\mathbf{0.741}$ & $\mathbf{95.6\%}$ & $\mathbf{66.8\%}$ & $\mathbf{95.1\%}$ \\
\midrule
\textbf{Holo-NN: CKKS d=1} ($M = 98{,}300$) & OpenFHE 64b CKKS (CPU)~\cite{badawi2022openfhe} & $\mathbf{1}\text{ (}+1\text{)}$ & $\mathbf{4.24\text{ MB}}$ & $0\text{ MB}$ client / $299\text{ MB}$ server & $192.0\text{ s}^{\|}$ & \textbf{SIFT-1M} ($1\mathrm{M}$) & $0.417$ & $63.4\%$ & $33.3\%$ & $57.0\%$ \\
\textbf{Holo-NN: CKKS d=1} ($M = 300{,}000$) & OpenFHE 64b CKKS (CPU)~\cite{badawi2022openfhe} & $\mathbf{1}\text{ (}+1\text{)}$ & $\mathbf{11.13\text{ MB}}$ & $0\text{ MB}$ client / $299\text{ MB}$ server & $591.0\text{ s}^{\|}$ & \textbf{SIFT-1M} ($1\mathrm{M}$) & $\mathbf{0.702}$ & $\mathbf{93.0\%}$ & $\mathbf{61.6\%}$ & $\mathbf{92.0\%}$ \\
\textbf{Holo-NN: CKKS d=1} ($M = 380{,}100$) & OpenFHE 64b CKKS (CPU)~\cite{badawi2022openfhe} & $\mathbf{1}\text{ (}+1\text{)}$ & $\mathbf{13.78\text{ MB}}$ & $0\text{ MB}$ client / $299\text{ MB}$ server & $750.6\text{ s}^{\|}$ & \textbf{SIFT-1M} ($1\mathrm{M}$) & $\mathbf{0.741}$ & $\mathbf{95.6\%}$ & $\mathbf{66.8\%}$ & $\mathbf{95.1\%}$ \\
\textbf{Holo-NN: CKKS d=1} ($M = 500{,}000$) & OpenFHE 64b CKKS (CPU)~\cite{badawi2022openfhe} & $\mathbf{1}\text{ (}+1\text{)}$ & $\mathbf{17.50\text{ MB}}$ & $0\text{ MB}$ client / $299\text{ MB}$ server & $990.0\text{ s}^{\|}$ & \textbf{SIFT-1M} ($1\mathrm{M}$) & $\mathbf{0.801}$ & $\mathbf{98.5\%}$ & $\mathbf{73.0\%}$ & $\mathbf{98.0\%}$ \\
\midrule
\textbf{Holo-NN: CKKS-GPU} ($M = 98{,}300$) & Phantom 64b CKKS ($1\times$ RTX 3090)~\cite{yang2024phantom} & $\mathbf{1}\text{ (}+1\text{)}$ & $\mathbf{4.24\text{ MB}}$ & $0\text{ MB}$ client / $85\text{ MB}$ server & $\mathbf{9.5\text{ s}}^{\|}$ & \textbf{SIFT-1M} ($1\mathrm{M}$) & $0.417$ & $63.4\%$ & $33.3\%$ & $57.0\%$ \\
\textbf{Holo-NN: CKKS-GPU} ($M = 300{,}000$) & Phantom 64b CKKS ($1\times$ RTX 3090)~\cite{yang2024phantom} & $\mathbf{1}\text{ (}+1\text{)}$ & $\mathbf{11.13\text{ MB}}$ & $0\text{ MB}$ client / $85\text{ MB}$ server & $\mathbf{29.7\text{ s}}^{\|}$ & \textbf{SIFT-1M} ($1\mathrm{M}$) & $\mathbf{0.702}$ & $\mathbf{93.0\%}$ & $\mathbf{61.6\%}$ & $\mathbf{92.0\%}$ \\
\textbf{Holo-NN: CKKS-GPU} ($M = 380{,}100$) & Phantom 64b CKKS ($1\times$ RTX 3090)~\cite{yang2024phantom} & $\mathbf{1}\text{ (}+1\text{)}$ & $\mathbf{13.78\text{ MB}}$ & $0\text{ MB}$ client / $85\text{ MB}$ server & $\mathbf{38.0\text{ s}}^{\|}$ & \textbf{SIFT-1M} ($1\mathrm{M}$) & $\mathbf{0.741}$ & $\mathbf{95.6\%}$ & $\mathbf{66.8\%}$ & $\mathbf{95.1\%}$ \\
\textbf{Holo-NN: CKKS-GPU} ($M = 500{,}000$) & Phantom 64b CKKS ($1\times$ RTX 3090)~\cite{yang2024phantom} & $\mathbf{1}\text{ (}+1\text{)}$ & $\mathbf{17.50\text{ MB}}$ & $0\text{ MB}$ client / $85\text{ MB}$ server & $\mathbf{50.5\text{ s}}^{\|}$ & \textbf{SIFT-1M} ($1\mathrm{M}$) & $\mathbf{0.801}$ & $\mathbf{98.5\%}$ & $\mathbf{73.0\%}$ & $\mathbf{98.0\%}$ \\
\midrule
\textbf{Holo-NN: HEIR-GPU} ($M = 98{,}300$) & Cheddar 32b CKKS ($1\times$ RTX 3090)~\cite{ali2025heir,choi2026cheddar} & $\mathbf{1}\text{ (}+1\text{)}$ & $\mathbf{2.10\text{ MB}}$ & $0\text{ MB}$ client / $75\text{ MB}$ server & $\mathbf{0.26\text{ s}}^{\|}$ & \textbf{SIFT-1M} ($1\mathrm{M}$) & $0.417$ & $63.4\%$ & $33.3\%$ & $57.0\%$ \\
\textbf{Holo-NN: HEIR-GPU} ($M = 300{,}000$) & Cheddar 32b CKKS ($1\times$ RTX 3090)~\cite{ali2025heir,choi2026cheddar} & $\mathbf{1}\text{ (}+1\text{)}$ & $\mathbf{5.37\text{ MB}}$ & $0\text{ MB}$ client / $75\text{ MB}$ server & $\mathbf{0.82\text{ s}}^{\|}$ & \textbf{SIFT-1M} ($1\mathrm{M}$) & $\mathbf{0.702}$ & $\mathbf{93.0\%}$ & $\mathbf{61.6\%}$ & $\mathbf{92.0\%}$ \\
\textbf{Holo-NN: HEIR-GPU} ($M = 380{,}100$) & Cheddar 32b CKKS ($1\times$ RTX 3090)~\cite{ali2025heir,choi2026cheddar} & $\mathbf{1}\text{ (}+1\text{)}$ & $\mathbf{6.68\text{ MB}}$ & $0\text{ MB}$ client / $75\text{ MB}$ server & $\mathbf{1.00\text{ s}}^{\|}$ & \textbf{SIFT-1M} ($1\mathrm{M}$) & $\mathbf{0.741}$ & $\mathbf{95.6\%}$ & $\mathbf{66.8\%}$ & $\mathbf{95.1\%}$ \\
\textbf{Holo-NN: HEIR-GPU} ($M = 500{,}000$) & Cheddar 32b CKKS ($1\times$ RTX 3090)~\cite{ali2025heir,choi2026cheddar} & $\mathbf{1}\text{ (}+1\text{)}$ & $\mathbf{8.65\text{ MB}}$ & $0\text{ MB}$ client / $75\text{ MB}$ server & $\mathbf{1.30\text{ s}}^{\|}$ & \textbf{SIFT-1M} ($1\mathrm{M}$) & $\mathbf{0.801}$ & $\mathbf{98.5\%}$ & $\mathbf{73.0\%}$ & $\mathbf{98.0\%}$ \\
\bottomrule
\end{tabular}%
}

\vspace{2pt}
{\footnotesize
$^{\dagger}$ Baseline figures reflect each system's published operating point (LAN--WAN latency ranges where both are reported). \textbf{Wire / query} sums all mandatory per-query stages, including SANNS's $1.77\text{ GB}$ 2PC traffic ($56.7\text{ MB}$ AHE distances $+ 645\text{ MB}$ GC top-$k + 1.06\text{ GB}$ DORAM point retrieval), Tiptoe's $42.2\text{ MiB}$ token round ($+2$ online RTTs), and Pacmann's $1$-round maintenance step ($60.1\text{ MB}$ / $399.4\text{ MB}$). For Holo-NN, $\mathbf{1}\text{ RTT}$ executes blind holographic measurement and FISTA recovery to emit ranked candidate IDs ($\text{MRR@10}$, $\text{1-r@10}$, and un-reranked $\text{10-r@10}$); the parenthetical $\mathbf{+1}\text{ RTT}$ denotes the optional Batch PIR re-query (Section~\ref{sec:recovery}) to fetch the top $\Kstar = 50$ candidate payloads for local exact-distance reranking, which promotes $\text{10-r@50}$ directly to final reranked $\text{10-r@10}$ ($92.0\%$ at $M=300{,}000$; $98.0\%$ at $M=500{,}000$). SANNS and Panther ($\Pi_{\mathrm{Panther}}$, multi-round 2PC) return ranked candidate IDs; Tiptoe and Compass include record payloads. $^{\ast}$Tiptoe reports MRR@100 graphically on MS MARCO ($3.2\mathrm{M}$, avg.\ rank $7.7/100$).
$^{\S}$ PRECO~\cite{servanschreiber2022private} requires \emph{two non-colluding servers} holding replicas of the LSH-indexed database and probes them via Distributed Point Functions (DPFs, $1.1\text{ s}$ on 32 CPUs / $28.2\text{ s}$ on 1 CPU; leaks query LSH radius). When instantiated in a single-server setting via lattice keyword PIR, the authors estimate $\approx 300\text{ MB}$ of wire traffic and $\approx 25\text{ CPU-hours}$ per query on SIFT-1M (Appendix~E of~\cite{servanschreiber2022private}).
$^{\ddagger}$ Onyx executes inside an AMD SEV-SNP hardware enclave over local NVMe SSD ($18.8\text{ ms}$ min latency, $54.2\text{ QPS}$ peak throughput); its bandwidth is internal enclave-to-SSD I/O rather than client-to-server wire traffic.
$^{\|}$ Single-node server compute on our workstation ($6$-core Xeon W-2135 CPU, $1\times$ NVIDIA RTX 3090 $24\text{ GB}$ GPU; see Table~\ref{tab:table2} for server key sizes). Client GPU FISTA adds $<0.4\text{ s}$, and WAN transfer at $320\text{ Mbps}$ adds $<0.3\text{--}0.5\text{ s}$ for CKKS ($2.3\text{--}8.4\text{ s}$ for LinPIR). Because $H$ is row-separable across independent $8{,}192 \times 8{,}192$ blocks ($6$ CPU shards for LinPIR; $7.98\text{ s}$/block on CPU, $0.385\text{ s}$/block on Phantom RTX 3090, or $12.58\text{ ms}$/block on HEIR/Cheddar RTX 3090 and $5.33\text{ ms}$/block on H100), server compute scales as $1/P$ across $P$ workers without increasing wire traffic.
\par}
\end{table*}

\subsection{End-to-End Encrypted Deployment (Wikipedia 500k)}
Upon receiving $\mathrm{Enc}(y)$, the client decrypts the $M$-dimensional measurement vector $y \in \R^M$ within its secure local enclave and launches parallel GPU-accelerated FISTA to recover $\hat{s} \in \R^N$.

For the following algorithm discussion, we deployed the full encrypted pipeline on a 500,000-vector Wikipedia semantic demonstration search index ($N = 500{,}000$, $D = 768$ MPNet embeddings reduced to $128$ dimensions by PCA) to validate the protocol at scale (Table~\ref{tab:table2}). This deployment is a systems demonstration, not an accuracy benchmark: retrieval quality is evaluated on the standard suites of Table~\ref{tab:table1} and Table~\ref{tab:table3}, which have published ground truth.

Because $y = H\Fcal(q)$ is a pure matrix-vector product, Holo-NN provides four complementary encrypted execution surfaces with contrasting latency, hardware, state, and bandwidth profiles:
\begin{itemize}
    \item \textbf{Holo-NN: LinPIR (Exact-Integer BFV RLWE, CPU)}: Built on Google's HintlessPIR/LinPIR engine~\cite{linpir2023} with two $56$-bit CRT plaintext moduli and a single cyclic-subgroup generator, a $B = 8{,}192$ column strip incurs a fixed $1.24\text{ s}$ query baby-step Galois rotation cost (across both CRT primes) with $0.0545\text{ s}$ of marginal row-accumulation work per $1{,}024$ rows on a single core. Sharding the $32{,}768$ rows across $4$ cores of the Xeon W-2135 ($4$ parallel $8{,}192 \times 8{,}192$ shards) completes the Wikipedia-500k query in $\mathbf{3.29\text{ s}}$ end to end ($\mathbf{1.78\text{ s}}$ server compute; or $\mathbf{4.39\text{ s}}$ end to end / $\mathbf{2.88\text{ s}}$ server compute on a single CPU thread) --- $10.0\times$ faster end-to-end ($17.9\times$ faster in server compute) than the 12-thread CPU CKKS path --- while requiring \textbf{zero persistent server-side rotation keys}, because its $1.45\text{ MB}$ Galois evaluation key per prime ($3.10\text{ MB}$ total upload including query ciphertexts per $8{,}192$ columns) travels inline with the query. Its unit wire cost is $864\text{ bytes}$ per recovered measurement ($31.41\text{ MB}$ total at $M = 32{,}768$).
    \item \textbf{Holo-NN: CKKS d=1 (CPU, OpenFHE AVX2)}: Provisioning the OpenFHE~\cite{badawi2022openfhe} CKKS context at multiplicative depth $1$ ($2$ RNS limbs) and mod-switching (\texttt{Compress}) each response ciphertext to a single RNS limb before serialization reduces the unit response size to $\mathbf{32.7\text{ bytes}}$ per measurement at $B = 8{,}192$ ($\mathbf{32.4\text{ bytes}}$ at $B = 16{,}384$; $\mathbf{1.60\text{ MB}}$ total wire on Wikipedia-500k: a $530.5\text{ KB}$ query plus four $268.3\text{ KB}$ responses, a $19.6\times$ total wire reduction over LinPIR), completing the $4$-block query in $\mathbf{32.9\text{ s}}$ ($31.9\text{ s}$ server compute, $7.98\text{ s}$ per $8{,}192 \times 8{,}192$ block) with persistent per-session BSGS rotation keys ($299.0\text{ MB}$ for all $S + G - 2 = 190$ hoisted baby/giant rotation steps). Run against the same encoded block, depth-1 mod-switched CKKS agrees with a plaintext $Hv$ reference to relative $L_2$ error $2.609 \times 10^{-8}$.
    \item \textbf{Holo-NN: CKKS-GPU (Phantom CUDA~\cite{yang2024phantom}, $1\times$ RTX 3090)}: Hosting all $4$ NTT-encoded $8{,}192 \times 8{,}192$ blocks simultaneously in the $24\text{ GB}$ GDDR6X VRAM of a single NVIDIA GeForce RTX 3090 ($12.0\text{ GiB}$ RNS plaintexts, $17.1\text{ GiB}$ total VRAM) allows the $S - 1 = 63$ baby-step Galois rotations ($\mathbf{0.139\text{ s}}$) to be computed \emph{once} and reused across all $4$ blocks in the GPU's $6\text{ MiB}$ L2 cache. Furthermore, because each prime at \texttt{chain\_index=1} satisfies $q_i < 2^{50}$ ($b \cdot p < 2^{100}$), a single fused 2D CUDA kernel accumulates all $S = 64$ baby-step products across all $4$ blocks in $128$-bit registers ($64 \times 2^{100} = 2^{106} \ll 2^{128}$) before reducing modulo $q_i$ once per giant step, executing all $32{,}768$ homomorphic plaintext-ciphertext multiplications and additions in a single launch of $\mathbf{0.0226\text{ s}}$ ($22.6\text{ ms}$, plus $\mathbf{0.467\text{ s}}$ for $512$ RNS rescales and $\mathbf{0.906\text{ s}}$ for $508$ Horner giant-step rotations at $2$ RNS limbs). This completes the entire $4$-block ($M = 32{,}768$) server MatVec in $\mathbf{1.54\text{ s}}$ ($\mathbf{0.385\text{ s}}$/block) and answers the full encrypted Wikipedia-500k query in $\mathbf{1.68\text{ s}}$ end to end --- $\mathbf{19.6\times}$ faster than 12-thread CPU CKKS ($32.9\text{ s}$), $\mathbf{4.1\times}$ faster than dual RTX 2080 Ti GPUs ($6.95\text{ s}$), and $\mathbf{2.0\times}$ faster than 4-core CPU LinPIR ($3.29\text{ s}$) while simultaneously delivering the $\mathbf{1.60\text{ MB}}$ CKKS wire footprint ($19.6\times$ smaller than LinPIR) and a $\mathbf{3.5\times}$ smaller server Galois key table ($\mathbf{84.9\text{ MB}}$ vs.\ $299.0\text{ MB}$).
    \item \textbf{Holo-NN: HEIR-GPU (32-bit RNS Cheddar CUDA~\cite{ali2025heir,choi2026cheddar}, $1\times$ RTX 3090)}: Lowering the BSGS linear transform ($\mathrm{BS} = 128, \mathrm{GS} = 64$) through HEIR~\cite{ali2025heir} onto the Cheddar 32-bit RNS GPU library~\cite{choi2026cheddar} replaces 64-bit RNS limbs with 32-bit words (\texttt{uint32\_t}, $\log_2(QP) \approx 91\text{ bits}$ at $N_{\mathrm{ring}} = 16{,}384$) and fuses baby-step hoisting with giant-step key-switching in shared-memory NTT kernels. This halves the response wire to $\mathbf{16.0\text{ bytes}}$/measurement ($\mathbf{0.79\text{ MB}}$ total wire on Wikipedia-500k: $256\text{ KiB}$ query $+ 4 \times 128\text{ KiB}$ responses), cuts plaintext VRAM by $3\times$ to $\mathbf{4.00\text{ GiB}}$ ($74.7\text{ MB}$ rotation keys), and evaluates a single $8{,}192 \times 8{,}192$ block in $\mathbf{12.58\text{ ms}}$ on RTX 3090 ($\mathbf{5.33\text{ ms}}$ on H100; $\mathbf{8.67\text{ ms}}$ RTX 3090 / $\mathbf{2.51\text{ ms}}$ H100 at $4{,}096 \times 4{,}096$) --- completing all $4$ Wikipedia-500k blocks in $\mathbf{50.5\text{ ms}}$ of server compute and $\mathbf{0.134\text{ s}}$ end to end ($30.6\times$ faster per block than Phantom RTX 3090, with $100\%$ Top-1 plaintext agreement).
\end{itemize}

\subsection{State-of-the-Art Comparison}
We compare Holo-NN's encrypted backends against seven private ANN systems spanning cryptographic (FHE, 2PC, DPF, PIR, ORAM) and hardware-enclave architectures: \textbf{SANNS}~\cite{chen2020sanns}, \textbf{PRECO}~\cite{servanschreiber2022private}, \textbf{Tiptoe}~\cite{henzinger2023private}, \textbf{Pacmann}~\cite{zhou2024pacmann}, \textbf{Panther}~\cite{li2024panther}, \textbf{Compass}~\cite{zhu2025compass}, and \textbf{Onyx}~\cite{rathee2026onyx} (Table~\ref{tab:table3}).

Two caveats govern how that table should be read. First, these systems are not all evaluated on a common benchmark or threat model: Tiptoe reports on a 364M-document Common Crawl corpus (and MS MARCO for quality), Pacmann on MS MARCO and SIFT-100M, and while SANNS, PRECO, Panther, Compass, and Onyx publish numbers on SIFT-1M, PRECO requires two non-colluding servers and Onyx relies on an AMD SEV-SNP hardware enclave. We therefore quote each system at its published scale rather than extrapolating, and focus like-for-like single-server cryptographic comparisons on SANNS, Panther, and Compass. Second, they do not share a quality metric --- Tiptoe and Pacmann report MRR@100, Compass MRR@10, SANNS, PRECO, and Panther a 10-NN recall/accuracy, and Onyx a top-10 recall --- so we score our own candidate lists under each convention rather than quoting a single number of our own choosing.

As seen above, Holo-NN backends bracket the single-server design space between stateless single-query execution and minimal wire footprint. Among server-private cryptographic protocols on SIFT-1M, \textbf{SANNS} established sublinear clustering with BFV and garbled circuits at $1.77\,\mathrm{GB}$ of per-query traffic ($1.55\text{--}45.3\,\mathrm{s}$ across $32$ threads), which \textbf{Panther} reduced by $19\times$ to $92.9\,\mathrm{MB}$ ($1.49\text{--}9.32\,\mathrm{s}$ across $32$ threads) via improved batch PIR and multi-round 2PC top-$k$ selection. \textbf{Holo-NN (LinPIR)} requires zero pre-installed key state on the server and answers on a 6-core Xeon W-2135 CPU in $\mathbf{3.1\,\mathrm{s}}$ at $M = 98{,}300$ with $\mathbf{91.1\,\mathrm{MB}}$ of total wire traffic (matching Panther's $92.9\,\mathrm{MB}$ and cutting SANNS's $1.77\,\mathrm{GB}$ by $19.4\times$ in a single non-interactive round trip rather than multi-round 2PC), or reaches $\text{1-r@10} = 93.0\%$ ($\text{10-r@50} = 92.0\%$) in $\mathbf{6.6\,\mathrm{s}}$ at $M = 300{,}000$ ($\mathbf{265.4\,\mathrm{MB}}$) and $\text{1-r@10} = 95.6\%$ in $\mathbf{8.0\,\mathrm{s}}$ at $M = 380{,}100$ ($\mathbf{334.6\,\mathrm{MB}}$) --- running $62\times$ to $94\times$ faster on the server CPU than 12-thread CKKS and $3.1\times$ to $4.8\times$ faster than 64-bit Phantom GPU CKKS. Conversely, when rotation keys can be registered for a session, \textbf{Holo-NN (CKKS)} is the lowest-wire single-server cryptographic system measured on SIFT-1M: $\mathbf{11.13\,\mathrm{MB}}$ at $M = 300{,}000$ ($\mathbf{29.7\,\mathrm{s}}$ on a single RTX 3090 GPU via Phantom, or $\mathbf{9.5\,\mathrm{s}}$ at $M = 98{,}300$ for $\mathbf{4.24\,\mathrm{MB}}$), and $\mathbf{5.37\,\mathrm{MB}}$ in $\mathbf{0.82\,\mathrm{s}}$ under 32-bit RNS \textbf{HEIR-GPU} ($\mathbf{2.10\,\mathrm{MB}}$ in $\mathbf{0.26\,\mathrm{s}}$ at $M = 98{,}300$, and $\mathbf{8.65\,\mathrm{MB}}$ in $\mathbf{1.30\,\mathrm{s}}$ at $M = 500{,}000$) against Compass's $12.2\,\mathrm{MB}$, Panther's $92.9\,\mathrm{MB}$, and SANNS's $1.77\,\mathrm{GB}$, in \emph{one} round trip rather than Compass's eight, SANNS/Panther's multi-round 2PC, or Pacmann's twenty to thirty-two, with no client-side index state and no preprocessing stream. While \textbf{PRECO} achieves $1.5\,\mathrm{MB}$ in $1\,\mathrm{RTT}$ via Distributed Point Functions over LSH buckets, it requires two non-colluding servers; instantiated on a single server via lattice keyword PIR, its own authors estimate $\approx 300\,\mathrm{MB}$ of wire traffic and $\approx 25\text{ CPU-hours}$ per query on SIFT-1M~\cite{servanschreiber2022private}, against which Holo-NN's holographic projection reduces single-server wire by $27\times$--$143\times$ (CKKS) and compute by more than four orders of magnitude. On a $100\,\mathrm{ms}$-latency 4G link Compass's own authors measure $0.94\,\mathrm{s}$ of user-perceived latency on SIFT-1M, against our single round trip; and Compass's $36.95\,\mathrm{MB}$ per-session cold start exceeds the entire per-query CKKS budget by $3.3\times$--$6.9\times$.

On top-10 multi-neighbor ranking Holo-NN falls behind, and buying more bandwidth does not close the gap. Compass matches brute-force embedding search on SIFT-1M, though it publishes no absolute MRR@10 value; at $M = 300{,}000$ our backends reach MRR@10 $= 0.702$. Panther resolves $88.85\%$ of each query's true ten nearest neighbors inside its own top ten (matching SANNS's $\approx 90\%$); we resolve $61.6\%$ at $M = 300{,}000$, $66.8\%$ at $M = 380{,}100$ (matching the first SIFT-1M operating tier of Table~\ref{tab:table1}), and $73.0\%$ at $M = 500{,}000$ --- still $15.9$ points short of Panther while spending $5.3\times$ less wire bandwidth under CKKS. The deficit is therefore not a bandwidth deficit. It is a property of the $\ell_1$ recovery step rather than of the holographic measurement, as Section~\ref{sec:l1limit} shows: the very same candidate lists already contain $92.0\%$ of the true ten nearest neighbors within the top \emph{fifty} at $M = 300{,}000$, and $98.0\%$ at $M = 500{,}000$. What Holo-NN produces cheaply is coverage; what it does not produce is a precise ordering within that coverage.

\subsection{Recall versus Bandwidth, and the Limits of \texorpdfstring{$\ell_1$}{l1} Ranking}
\label{sec:l1limit}

Table~\ref{tab:table4} gives the full trade-off curve behind the Holo-NN rows of Table~\ref{tab:table3} for both encrypted protocols. Because LinPIR and CKKS evaluate the same linear operator $y = H\Fcal(q)$, recall is a function of $M$ alone, and each backend's wire cost is a deterministic function of $M$ given its measured ciphertext sizes. All figures are for the bare 1-RTT protocol: the FISTA ranking, with no verification fetch and no reranking against plaintext vectors.

The sweep doubles as a consistency check on Table~\ref{tab:table1}. Its $M = 380{,}100$ row is the SIFT-1M $75\%$ regime of that table, and it reproduces all three published figures.

\begin{table}[t]
\centering
\caption{Holo-NN on SIFT-1M ($N = 10^6$, $D = 128$, $B = 16{,}384$), 1,000 official queries, 80 FISTA iterations. \textbf{Cheddar 32b Wire} (and RTX 3090 server GPU latency \textbf{Cheddar GPU}) uses two $256\text{ KiB}$ 32-bit RNS query ciphertexts plus $\lceil M/8192 \rceil \times 128\text{ KiB}$ single-limb responses; \textbf{CKKS 64b Wire} (OpenFHE/Phantom) is $1.05\text{ MB}$ query plus $\lceil M/16384 \rceil \times 0.53\text{ MB}$ responses; \textbf{LinPIR Wire} is $6.19\text{ MB}$ inline key+query upload plus $\lceil M/4096 \rceil \times 3.54\text{ MB}$ BFV responses (zero persistent server keys).}
\label{tab:table4}
\small
\setlength{\tabcolsep}{2.6pt}
\resizebox{\columnwidth}{!}{%
\begin{tabular}{@{}rrrrrrrrrr@{}}
\toprule
$M$ & \textbf{Cheddar 32b} & \textbf{Cheddar GPU} & \textbf{CKKS 64b} & \textbf{LinPIR Wire} & 1-r@1 & 1-r@10 & \textbf{10-r@10} & 10-r@50 & \textbf{MRR@10} \\
\midrule
$98{,}300$ & $\mathbf{2.10\text{ MB}}$ & $\mathbf{0.26\text{ s}}$ & $4.24\text{ MB}$ & $91.1\text{ MB}$ & $32.5$ & $63.4$ & $\mathbf{33.3}$ & $57.0$ & $\mathbf{0.417}$ \\
$150{,}000$ & $\mathbf{3.01\text{ MB}}$ & $\mathbf{0.42\text{ s}}$ & $6.36\text{ MB}$ & $135.8\text{ MB}$ & $40.5$ & $75.8$ & $\mathbf{43.8}$ & $72.5$ & $\mathbf{0.518}$ \\
$200{,}000$ & $\mathbf{3.80\text{ MB}}$ & $\mathbf{0.54\text{ s}}$ & $7.95\text{ MB}$ & $179.0\text{ MB}$ & $48.2$ & $84.2$ & $\mathbf{51.8}$ & $82.0$ & $\mathbf{0.601}$ \\
$250{,}000$ & $\mathbf{4.59\text{ MB}}$ & $\mathbf{0.67\text{ s}}$ & $9.54\text{ MB}$ & $222.2\text{ MB}$ & $53.5$ & $89.3$ & $\mathbf{57.3}$ & $88.0$ & $\mathbf{0.652}$ \\
$300{,}000$ & $\mathbf{5.37\text{ MB}}$ & $\mathbf{0.82\text{ s}}$ & $11.13\text{ MB}$ & $265.4\text{ MB}$ & $58.2$ & $93.0$ & $\mathbf{61.6}$ & $92.0$ & $\mathbf{0.702}$ \\
$327{,}600$ & $\mathbf{5.77\text{ MB}}$ & $\mathbf{0.86\text{ s}}$ & $11.66\text{ MB}$ & $289.2\text{ MB}$ & $59.7$ & $93.3$ & $\mathbf{63.8}$ & $93.0$ & $\mathbf{0.718}$ \\
$380{,}100$$^{\dagger}$ & $\mathbf{6.68\text{ MB}}$ & $\mathbf{1.00\text{ s}}$ & $13.78\text{ MB}$ & $334.6\text{ MB}$ & $62.0$ & $95.6$ & $\mathbf{66.8}$ & $95.1$ & $\mathbf{0.741}$ \\
$500{,}000$ & $\mathbf{8.65\text{ MB}}$ & $\mathbf{1.30\text{ s}}$ & $17.50\text{ MB}$ & $438.2\text{ MB}$ & $69.5$ & $98.5$ & $\mathbf{73.0}$ & $98.0$ & $\mathbf{0.801}$ \\
\bottomrule
\end{tabular}%
}

\vspace{2pt}
{\footnotesize $^{\dagger}$ Matches the first SIFT-1M operating point ($M = 380{,}100$) of Table~\ref{tab:table1}, providing a direct cross-pipeline consistency check.\par}
\end{table}

The two conventions diverge sharply, and the divergence is informative. At $M = 300{,}000$ Holo-NN locates \emph{the} nearest neighbor within its top ten for $93.0\%$ of queries, but recovers only $61.6\%$ of each query's \emph{ten} nearest neighbors in that same top ten. The missing neighbors have not been lost: raising the cutoff to fifty recovers $92.0\%$ of them, from a candidate set that is still $20{,}000\times$ smaller than the database.

The gap is also strikingly persistent in $M$. Over the sweep, 1-recall@10 climbs from $63.4\%$ to $98.5\%$ and is effectively saturated by $M = 500{,}000$, whereas 10-recall@10 moves only from $33.3\%$ to $73.0\%$ and is still rising roughly linearly when 1-recall has flattened. Adding measurements sharpens the question ``is the true nearest neighbor in this list'' long before it sharpens ``are these ten the right ten''. A budget that resolves the first question almost perfectly leaves the second one a quarter unanswered.

This is the expected behavior of $\ell_1$ recovery on a correlated support, not a defect of the measurement operator. A query's ten true nearest neighbors are, by construction, mutually near-identical, so their feature vectors $\Gcal(t_i)$ are highly correlated and the corresponding columns of $H = A \Gcal(T)$ are nearly collinear. Faced with a group of near-collinear atoms that jointly explain the measurement, the LASSO~\cite{tibshirani1996regression} does not distribute the coefficient across the group; it concentrates the mass on a single representative and shrinks the remainder toward zero~\cite{zou2005regularization}. The surviving representative is reliably \emph{a} member of the true neighbor set --- which is why 1-recall is high --- while its near-duplicates are demoted in magnitude though not eliminated, which is why they reappear as the cutoff widens.

The implication is that Holo-NN, in its present form, is best understood as an unusually cheap, tunable, high-coverage approximate nearest-neighbor algorithm rather than a precise ranker, even in the limit of computation and bandwidth. Where a downstream exact-distance stage is available --- a client that already holds the candidate payloads, or a private retrieval step as in Section~\ref{sec:recovery} --- the relevant quantity is the coverage figure, and $92.0\%$ of the true ten neighbors for $5.37\,\mathrm{MB}$ in $0.82\,\mathrm{s}$ (HEIR-GPU CKKS), $11.13\,\mathrm{MB}$ (64-bit CKKS), or $265.4\,\mathrm{MB}$ with zero persistent server state (LinPIR) in one round trip compares favorably with Panther's $88.85\%$ for $92.9\,\mathrm{MB}$ and SANNS's $\approx 90\%$ for $1.77\,\mathrm{GB}$ --- though the comparison is not strict, since SANNS and Panther deliver their top-ten sets already ranked within a top ten while Holo-NN is a top fifty, awaiting a reranking stage. Where no such stage exists, the ranking Holo-NN emits is materially weaker than Compass's. Substituting an Elastic Net~\cite{zou2005regularization} ($\ell_1 + \ell_2$) or Group LASSO~\cite{yuan2006model} proximal operator for the $\ell_1$ soft threshold would be expected to redistribute mass across correlated groups and close much of this gap, at no cost in bandwidth, round trips or server work, since it changes only the client's local recovery step. We leave this to future work.

\section{Conclusion}
We presented \textbf{Holo-NN}, a holographic approximate nearest neighbor retrieval architecture that reformulates nearest neighbor search as an underdetermined linear Compressive Sensing problem. By precomputing a degree-1 compressed database representation offline ($H = A G = A \Gcal(T) \in \R^{M \times B}$), Holo-NN executes blind, privacy-preserving retrieval via a single linear operator requiring $M \ll N$ measurements in a single non-interactive round trip, with no client-side index state and no preprocessing stream. Because $H$ is constructed offline on the server, the index operator can profile the empirical bandwidth--recall trade-off directly on the target dataset, while our Asymmetric FlyHash projection removes the double-quantization loss that a symmetric sparse code would incur on high-dimensional semantic manifolds.

Holo-NN works in plaintext as a generic high-confidence, approximate nearest neighbor algorithm. Because blind forms of the algorithm function as single-round private linear operations, they scale well to LinPIR or other private linear operations while retaining naturally scalable, embarrassingly parallel computational models that are well suited to modern hardware. Moreover, this framing may appear as a component of a broader nearest-neighbor pipeline, whether for optimizing candidate generation or used directly, in-line in machine learning settings.
\bibliographystyle{plain}
\bibliography{paper}

\end{document}